\documentclass{emulateapj}
\usepackage{natbib}
\usepackage{amsmath}
\usepackage{appendix}
\usepackage[LGRgreek]{mathastext}
\usepackage{rotating}
\usepackage{longtable}

\newcommand{\sqarcsec}{$\mathrm{arcsec^{-2}}$ }
\newcommand{\ks}{$K_s$\,}
\newcommand{\wise}{\emph{WISE}\,}
\newcommand{\galex}{\emph{GALEX}\,}
\newcommand{\twomass}{\emph{2MASS}\,}
\newcommand{\fuvminusk}{$[FUV-K_s]$\,}
\newcommand{\micronm}{$\mu$m\,}
\newcommand{\wIIIminuswIV}{$[12\mu m-22\mu m]$\,}
\newcommand{\wIminuswII}{$[3.4\mu m-4.6\mu m]$\,}

\shorttitle{Constraints on Feedback in the Local Universe}
\shortauthors{Vaddi et al.}

\begin{document}
\title{Constraints on Feedback in the local Universe:  The relation between star formation and AGN activity in early type galaxies}

\author{Sravani Vaddi\altaffilmark{1}, Christopher P. O'Dea\altaffilmark{1,2}, Stefi A. Baum\altaffilmark{1,2}, Samantha Whitmore\altaffilmark{3}, Rabeea Ahmed\altaffilmark{3}, Katherine Pierce \altaffilmark{4}, Sara Leary \altaffilmark{1}}
\altaffiltext{1}{Rochester Institute of Technology, Rochester, NY 14623, USA}
\altaffiltext{2}{University of Manitoba,  Department of Physics and Astronomy, Manitoba, Canada}
\altaffiltext{3}{Harvard University Cambridge, MA 02138, USA}
\altaffiltext{4}{University at Buffalo, Buffalo, NY, 14260, USA}

\email{sravani.vaddi@gmail.com}
\begin{abstract}
We address the relation between star formation and AGN activity in a sample of 231 nearby ($0.0002<z<0.0358$) early type galaxies by carrying out a multi-wavelength study using archival observations in the UV, IR and radio.  Our results indicate that early type galaxies in the current epoch are rarely powerful AGNs, with $P<10^{22}\,WHz^{-1}$ for a majority of the galaxies.  Only massive galaxies are capable of hosting powerful radio sources while less massive galaxies are hosts to lower radio power sources.  Evidence of ongoing star formation is seen in approximately 7\% of the sample.  The SFR of these galaxies is less than 0.1 $M_{\odot}yr^{-1}$.  They also tend to be radio faint ($P<10^{22}\,WHz^{-1}$). There is a nearly equal fraction of star forming galaxies in radio faint ($P<10^{22}\,WHz^{-1}$) and radio bright galaxies ($P\geq10^{22}\,WHz^{-1}$) suggesting that both star formation and radio mode feedback are constrained to be very low in our sample. We notice that our galaxy sample and the Brightest Cluster Galaxies (BCGs) follow similar trends in radio power versus SFR. This may be produced if both radio power and SFR are related to stellar mass.

\end{abstract}


\section{Introduction}
	It is now well known that supermassive black holes (SBH) are present in the centers of massive galaxies \citep{kormendy1995} and share interesting correlations with the host galaxy properties such as the velocity dispersion \citep{ferrarese2000,gebhardt2000}, bulge mass \citep{haring2004}, bulge luminosity \citep{kormendy1995,magorrian1998} and galaxy light concentration \citep{graham2001}.  These empirical correlations suggest that the growth of the central SBH and the  host galaxy are fundamentally interlinked.  AGN feedback may be responsible for the correlations observed \citep{silkrees1998,king2003,fabian2012}, although it has also been argued that the origin of the observed relations is entirely non-causal and is a natural consequence of merger driven galaxy growth \citep{peng2007,jahnke2011,graham2013}.  The energy released from the central SBH is several orders more than the binding energy of massive galaxies \citep{fabian2012}.  This energy has the potential to expel gas from the  galaxy (radiative-mode feedback) or deposit energy into the surroundings  and thus heat up the inter galactic medium (mechanical feedback).  These two modes may operate at different redshifts and accretion rates and ensure to regulate the growth of the black hole and the galaxy \citep[review of ][]{mcnamara2007, fabian2012, churazov2005}.\\	
	
	Various theoretical models that invoke AGN feedback in galaxy evolution are also able to successfully reproduce the observed galaxy luminosity function \citep{silkrees1998,king2003,granato2004,dimatteo2005,springel2005,croton2006,hopkins2008}.   This theoretical picture has been supported by numerous observations.  The strongest evidence comes from the brightest cluster galaxies (BCG) of cool core clusters, whose powerful radio jets have swept out cavities in the intracluster medium  (ICM)\citep{rosner1989,allen2001,mcnamara2007}.  While in some individual galaxies, energy transportation into the ISM via AGN driven outflows are observed to remove gas from the central regions of the galaxy \citep{crenshaw2003,nesvadba2007,alexander2010,morganti2013}.  
All these show the negative effect of AGN feedback by removing/heating up the gas and eventually suppressing the star formation and regulating the galaxy growth.  However, several other theoretical studies \citep{begelman1989,rees1989,silk2005,santini2012, silk2013} have reported an increased star formation rate in AGN especially at high redshifts via induced pressure by jets/winds.  All this evidence thus far has been obtained mostly from studies of large groups, clusters and galaxies at higher redshifts.  At low redshifts, the majority of the luminous AGNs reside in early type galaxies \citep{mclure1999, bahcall1997}.  But how common AGN feedback is in the local universe is not yet well understood.


	To explore this, we focused our attention on a carefully selected sample of nearby  early type galaxies and studied them at multiple wavelengths.  We describe the sample selection in Section \ref{sample}.  In Section \ref{data}, we discuss the data, steps carried out to retrieve the magnitude in the UV and in the IR, flux from radio images and extinction correction. We present our results in Section \ref{results}.  In Section \ref{summary} we discuss the implications of the results.  In the Appendix, we describe the photometry technique in more detail.


\section{The sample}
  \label{sample}
This study is focused on a sample of early type (ellipticals and S0) galaxies that are present at low redshift.  The sample was selected from the \emph{Two Micron All Sky Survey} (\twomass; \citep{jarrett2003}) that have an apparent $K_{s}$ band (2.2 $\mu$m) magnitude of 13.5 and brighter and whose positions correlate with the Chandra archive of ACIS-I and ACIS-S observations (C. Jones, private communication).  A total of 231 galaxies were identified.  The Chandra selection criteria was used to create a sample which would allow a study of the nature of AGN activity in early type galaxies.  X-ray emission is detected for approximately 80\% of the galaxies.  The X-ray luminosities of the nuclei range from $10^{38}$ to $10^{41}$ erg $s^{-1}$.  The Eddington ratios are measured to be small $\sim$ $10^{-5}$ to $10^{-9}$ suggesting that these galaxies are low-luminosity AGNs \citep{jones2013}. In this paper, we present a parallel study on the star formation in the sample.  In a second paper in this series, we will study the relation between the X-ray properties and the star formation.  The data are homogeneous and since the sample is not selected based on specific properties in radio or UV, the data can be considered to be unbiased regarding their star formation and radio source properties.  Our large dataset at low redshift allows us to study the interplay between star formation and AGN activity in typical galaxies in the current epoch.  \\

\begin{figure}
  \centering
    \includegraphics[width=0.5\textwidth]{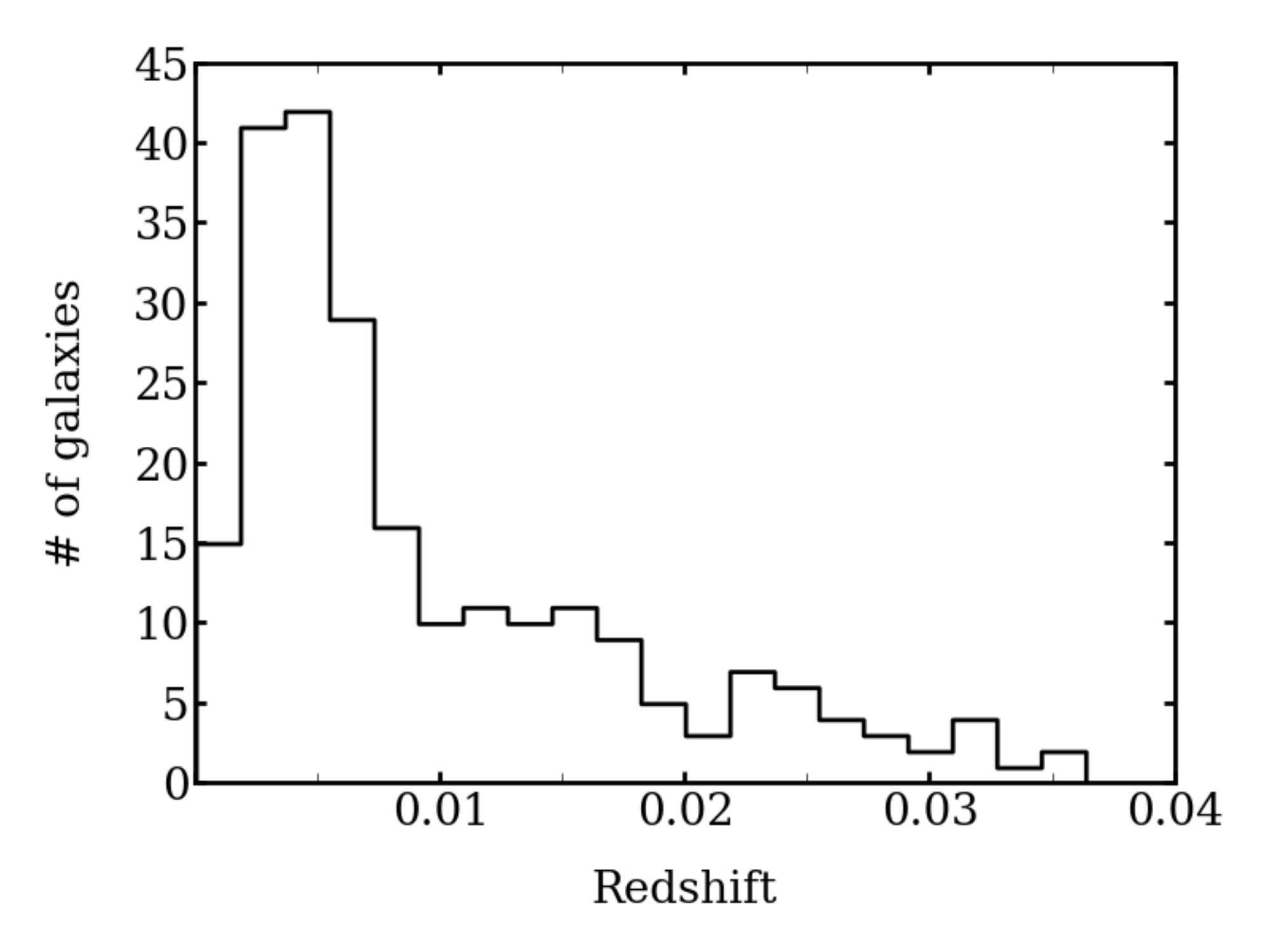}
    \caption{Figure shows the histogram of redshift for the sample indicating that most of the galaxies are at low redshifts.  Redshift of 0.01 corresponds to a distance of 42 Mpc.}
    \label{fig:redshift_hist}
  \end{figure}

Figure ~\ref{fig:redshift_hist} shows the redshift distribution of the sample.  All the galaxies are nearby galaxies.  The redshift range of the galaxies in the sample is $0.0002<z<0.0358$ with a median of $z=0.006$, of which, 63\% are at a redshift of less than 0.01.  Adopting a Hubble constant of $71 \,km\, s^{-1}Mpc^{-1}$ \citep{jarosik2011}, $z=0.01$ corresponds to a distance of 42 Mpc and 1\arcsec \; correspond to a scale of 210 pc.\\

\section{The Data}
\label{data}
We examine observations in multiple wavelengths for this study namely, radio, IR and UV.  Infrared observations were collected from \emph{Wide-field Infrared Survey Explorer} \wise\citep{wright2010} and \twomass \citep{jarrett2003}.  We use the \ks band to trace the stellar mass distribution;  \wise and \galex data to study star formation and radio data at 1.4 GHz  to study AGN properties.  Basic and observational properties for a subset of the  sample are given in Table ~\ref{tab:sample}.  A complete list of the sample properties in machine readable format can be obtained in the online version.\\

	\subsection{IR data}
The Two Micron All Sky Survey (\twomass) was conducted in the near-infrared $J(1.25 \mu m)$, $H(1.65 \mu m)$ and $K_{s}(2.16 \mu m)$ wavebands using two 1.3 m diameter telescopes with a resolution of $\sim$ 2\arcsec-3\arcsec.  The detectors are sensitive to point sources brighter than 1 mJy at the 10$\sigma$ level.  The astrometric accuracy is on order of 100 mas.  The camera contains three NICMOS 256$\times$256 HgCdTe arrays.  The \emph{WISE} mission observed the entire sky at four infrared wavebands - $W1$ at 3.4 $\mu$m, $W2$ at 4.6 $\mu$m, $W3$ at 12 $\mu$m, and $W4$ at 22 $\mu$m with an angular resolution of 6.1\arcsec, 6.4\arcsec, 6.5\arcsec and 12.0\arcsec respectively.  The field of view (FOV) is 47\arcmin.  The short wavelength detectors are HgCdTe arrays whereas long wavelength detectors are SiAs BIB arrays.  The arrays are 1024$\times$1024 pixels in size. WISE has 5$\sigma$ point source sensitivity higher than 0.08, 0.11, 1 and 6 mJy at 3.4, 4.6,12 and 22 $\mu m$ wavelengths respectively.\\
Rather than use the existing cataloged values, we chose to perform photometry on \twomass \ks band images and \wise band images for the following reasons:

\begin{enumerate}
	\item Underestimation of the WISE flux:  The extended source photometry for WISE is based on the \twomass aperture.  Since WISE is most sensitive to galaxies in W1 and W2 bands, the extended emission in these bands is much larger than the \twomass aperture. The \twomass aperture is too small by 10-20\% for resolved sources, thus resulting in an underestimate of the total flux by about 30-40\% ~\citep{cutri2012}. 

	\item Contamination of the flux from nearby sources:  In galaxies that have close companions, such as stars or other galaxies, there is an over estimation of the total flux since no masking was employed to remove the flux from unwanted sources.
\end{enumerate}

Surface photometry of our galaxy sample was performed using the ELLIPSE task in IRAF.   The task reads an input image and initial guess isophotal parameters and then returns the fitted isophote parameters and several other geometrical parameters.  A program has been written in Pyraf to automate the photometry for the entire sample.  The program runs the ELLIPSE task in two stages.  In the first stage,  the RA and DEC positions of \twomass as the initial central values were used and allowed for a non-linear increase of the semi-major axis step size.  In images where the target galaxy has close companions (e.g. saturated stars or or other galaxies), masks were created for those regions and the ELLIPSE task was allowed to flag the pixels in the mask.   The region masks were created by inspecting each galaxy by eye and is not done by the program itself.  These regions are then given as input to the program that calls the MSKREGIONS task.  This task creates pixel masks which in turn are used by the ELLIPSE task.\\

	In the first stage, the ellipse parameters are allowed to change freely.  From the output of the first stage, we extract the ellipse parameters (center, PA and ellipticity) of the isophote at which the intensity is 3$\sigma$ above the mean of the sky.  This gives the set of ellipse parameters that best describes the outer isophotes of the galaxy.  The mean and the standard deviation of the sky background is estimated by calculating the mode of an annulus region around the galaxy using the task FITSKY.  The task also allows for k-sigma clipping to reject any deviant pixels. \\

The program uses the new set of parameters obtained in the first stage as the initial guess for the second stage of the ELLIPSE run while keeping the center, PA and the ellipticity  fixed.  The task outputs a list of isophotes that have the same center, PA and ellipticity at different semi-major axes.  Ideally, total brightness of a galaxy is obtained by integrating the light from the entire galaxy.  Since galaxies do not have well defined edges and additionally our observations are limited by the sensitivity of the telescope, we integrate the light out to an isophote of specified brightness or to a specified radius.  For our analysis, the isophote whose intensity is one standard deviation above the mean of the sky background is considered the best aperture for that galaxy, encloseing all of the visible galaxy light.  Flux enclosed by this aperture is considered to be the $\lq$total flux' of the galaxy.  This process is repeated for all the galaxies in each band ($K_s,$ W1, W2, W3 and W4).\\

\begin{figure*}[!ht]
		\centering
		\includegraphics[width=0.7\textwidth]{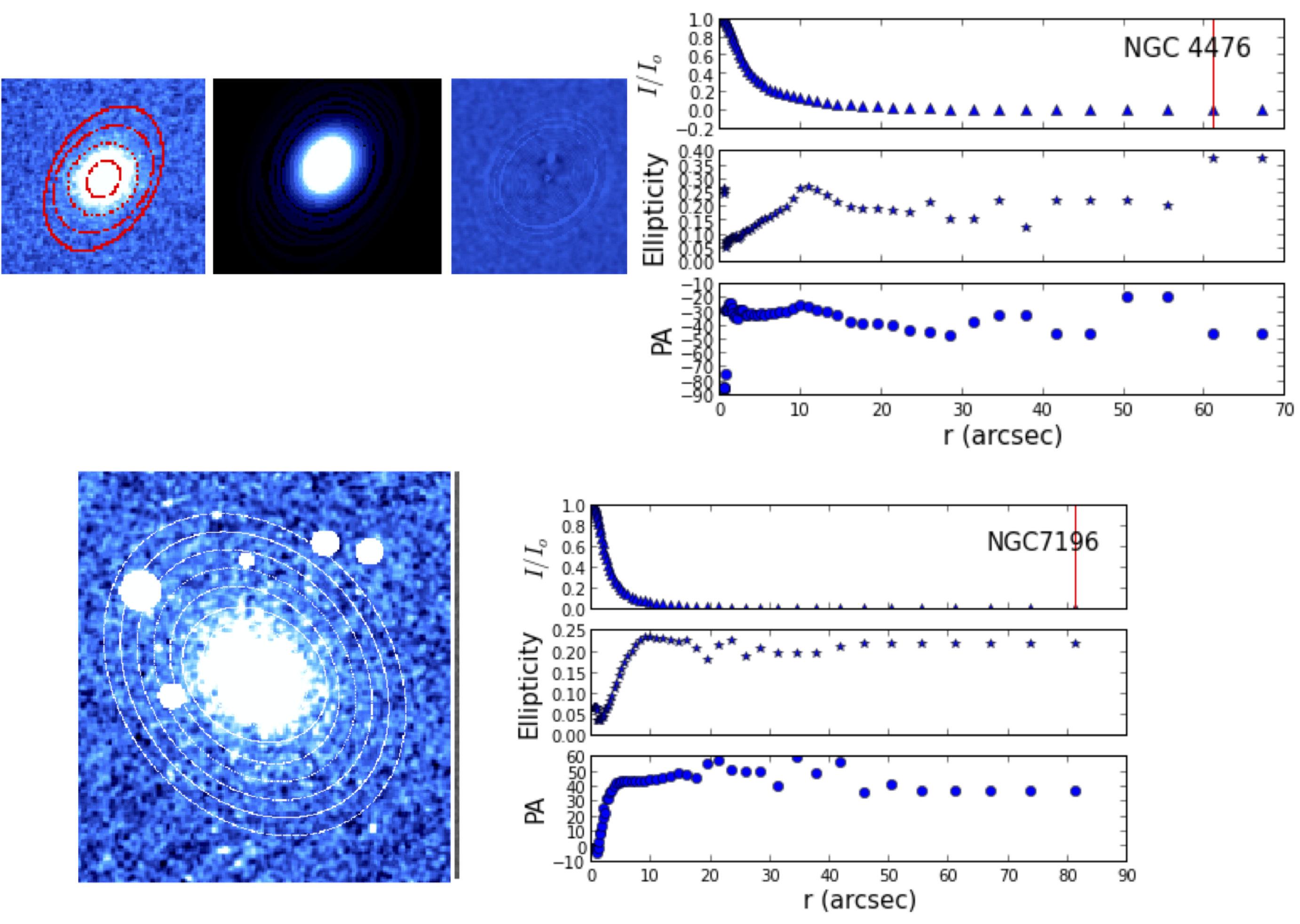}
		\caption{The top left panel shows a typical galaxy with ellipse fit, galaxy model and the residual image respectively.  The top right panel shows a graph of the normalized intensity, ellipticity and PA of the fit ellipses.  The ellipticity and PA of the ellipses are quite steady.  The red line is the semi-major axis where the intensity is one standard deviation above the sky.  The bottom left panel shows the ellipse fit to a galaxy for which neighboring bright objects have been masked.  The bottom right panel shows a graph of the normalized intensity, ellipticity and PA of the fit ellipses.}
		\label{fig:modelfit}
	\end{figure*}

A sample galaxy is shown in Figure ~\ref{fig:modelfit}.  The top panel shows the ellipse fit to a sample galaxy NGC 4476.  The middle panel shows the 2D smooth model image obtained from the isophotal analysis.  A residual image is obtained by subtracting the model from the galaxy image.  This is shown in the right panel.  The residual is nearly  smooth.    Also shown is a plot of the mean isophotal intensity with respect to the semi-major axis.  The mean intensity is normalized with the peak intensity.  The solid red line marks the semi-major axis at which the intensity drops to one $\sigma$ above the mean of the sky.  The ellipticity and the position angle as a function of the semi-major axis are also shown.  \\

Shown in the bottom panel of Figure ~\ref{fig:modelfit} is the ellipse fit to a sample galaxy in which masking of close companion has been done.  The ellipticity and the position angle are quite robust.

  \subsection{Radio data}
Radio data for our sample are drawn from several resources.  It is obtained from NRAO VLA Sky Survey (NVSS; \citep{condon1998})\footnote{http://www.cv.nrao.edu/nvss/NVSSlist.shtml}, Faint Images of the Radio Sky at Twenty-Centimeters (FIRST; \citep{becker1995})\footnote{http://sundog.stsci.edu/cgi-bin/searchfirst}, Sydney University Molonglo Sky Survey (SUMSS; \citep{mauch2003})\footnote{http://vizier.u-strasbg.fr/viz-bin/VizieR-3}.  We retrieved total flux density for the galaxies in the sample.\\

NVSS is a radio continuum survey conducted using Very Large Array (VLA) at 1.4 GHz and covers the entire sky north of $-40^{\circ}$ declination.  The array is used in D (with baseline of 1 km) and DnC (a hybrid configuration in which antennas in the east and the west arms are maintained in D configuration while the northern arm remain in C configuration with a baseline of 3.6 km) configuration and has angular resolution of 45\arcsec.  The catalog has 3$\sigma$ detection limit of $S\sim1.35$ mJy with typical RMS of 0.45 mJy/beam.  FIRST is the survey conducted over 10,000 square degrees of the North and South Galactic Caps.  The array is used in B-configuration with frequency centered at 1365 and 1435 MHz which gives a resolution of 5\arcsec.  The 5$\sigma$ sensitivity of the survey is $\sim$1 mJy.  The typical rms is 0.2 mJy.  If NVSS images are unavailable, FIRST images were used.  We estimated the total flux density using the TVWIN+IMSTAT tasks in AIPS.\\

\indent
For sources that lie in the southern hemisphere of the sky, data from SUMSS is used.  SUMSS is a survey carried out at 843 MHz with the Molonglo Observatory Synthesis Telescope (MOST).  The survey covers 8000 square degrees from $-30$ degrees declination southwards.  The RMS noise level is $\sim$1 mJy/beam with a detection limit of 5 mJy.  The SUMSS resolution of 43$\arcsec$ matches with NVSS resolution and thus, SUMSS and NVSS together provide a complete survey of the radio sky.   To match the 843 MHz flux density to 1.4 GHz flux density,  we use a simple linear extrapolation using the relation $S = S_{o}\nu^{-\alpha}$ with a power law index of 0.83 \citep{mauch2003}.  Hence we use the following relation to estimate the flux density at 1.4 GHz using the flux density at 843 MHz,
		\begin{equation}
			S_{1.4GHz} = S_{843MHz}\left(\frac{1400 MHz}{843 MHz}\right)^{-0.83}
		\end{equation}

The total radio power is calculated using the following equation
  \begin{equation}
  P_{1.4GHz} = S_{1.4GHz} 4 \pi D_L^2\;,
  \end{equation}
where $D_L$ is the luminosity distance calculated using Equation ~\eqref{eq:distance} in the Appendix.\\

 For sources that are not detected in any of the surveys, we searched in the VLA image archive.  We also searched the literature for the total flux density measurements.  If the radio flux density could not be determined using the above mentioned resources, the catalog detection limit at the source position is taken as an upper limit in the detection.

\subsection{UV data}
Far-UV (FUV) and Near-UV (NUV) magnitudes are collected from Galaxy Evolution Survey (GALEX; \citep{martin2005}).  GALEX is a wide-field UV imaging survey performed in two UV bands: FUV ($\lambda_{eff} = 1539\AA, \Delta\lambda=1344-1786\AA$) and NUV ($\lambda_{eff} = 2316\AA, \Delta\lambda=1771-2831\AA$) with angular resolution of 4.2\arcsec and 5.3\arcsec respectively.  The data is retrieved from GALEX GR6 release\footnote{http://galex.stsci.edu/GR6/?page=mastform}.  The  GALEX pipeline photometry estimates the magnitudes using the Kron-type SExtractor MAG\_AUTO aperture. It uses an elliptical aperture with the characteristic radius of the ellipse given by the first moment of the source brightness distribution.   The limiting magnitude for FUV and NUV is 24.7 and 25.5 respectively.\\

\subsection{Galactic Extinction Correction}

			FUV, NUV and \ks band magnitudes are corrected for galactic extinction using the corrections from \citep{wyder2005}.  For FUV, NUV and K bands, the ratio of $\frac{A(\lambda)}{E(B-V)}$ is 8.376, 8.741 and 0.347 where $A_{\lambda}$, the extinction at wavelength $\lambda$, is the difference between the observed magnitude and the actual magnitude of the source.  Values of color excess $E(B-V)$ are retrieved from the \galex GR6 catalog which uses the Schlegel maps for the reddening.  For galaxies whose $E(B-V)$ are unavailable in the GR6 catalog, we obtained it from NED (which also estimates reddening using Schlegel maps).  Because  the extinction for the infrared WISE bands is minimal, the correction for these bands have been ignored.\\ \\


\section{Results}
\label{results}

\subsection{Properties of star formation in the sample galaxies}
	\label{Radio-FUV-K}
	
\begin{figure}
	  \centering
	  \includegraphics[width=0.5\textwidth]{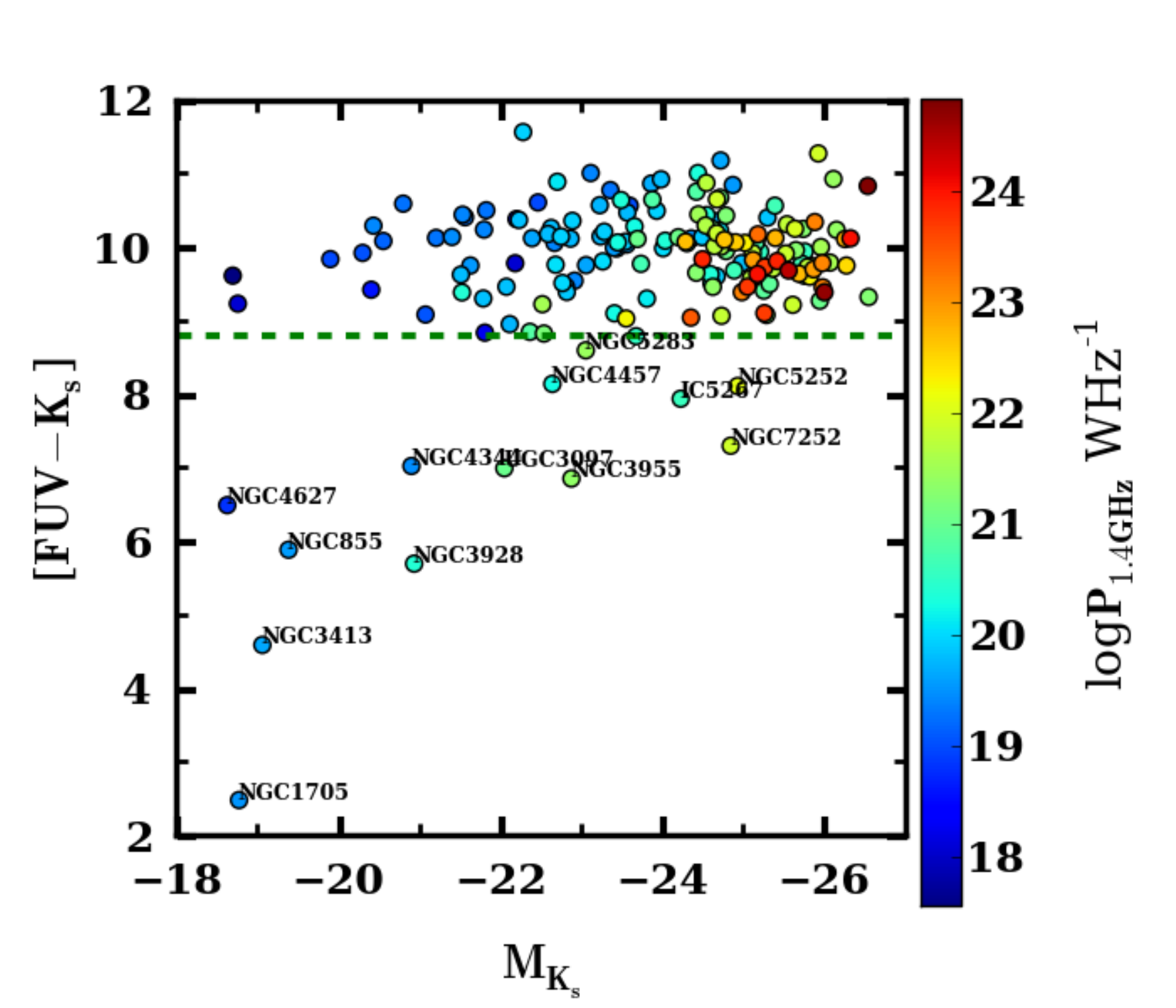}
	  \caption{\textbf{\fuvminusk versus absolute \ks magnitude}.  The green dashed line at 8.8 mag defines the separation between star forming and non-star forming as per \citep{gilde2007}.  The galaxies are color coded to represent the strength of the radio power.  Galaxies that are redder than 8.8 mag do not show signs of star formation; these consist of $\sim$92\% of the sample.  Galaxies that have \fuvminusk bluer than 8.8 mag show indications of significant star formation; they are less luminous (thus less massive) and are also weak in the radio.  The more massive galaxies tend to be FUV faint but are more luminous in the radio.  Note here that \ks magnitudes are in the Vega system and FUV are in the AB system.}
	  \label{fig:fuv-k_mk}
	\end{figure}

	 A color magnitude plot such as the one shown in Figure ~\ref{fig:fuv-k_mk}  can be used to distinguish between actively star-forming and  more passive galaxies.  The galaxies are color coded according to the strength of the radio power.  Those with high radio power are shown in red and low radio power galaxies are shown in blue.  We do not see a systematic change in the \fuvminusk color with absolute \ks magnitude.  An \fuvminusk color of 8.8 mag is defined  as the transition point between non-star-forming and star-forming galaxies \citep{gilde2007}.   Consistent with the color properties of typical early type galaxies, most of the galaxies in our sample lie in a band of \fuvminusk $\sim$ 9-11  \citep{gilde2007}.  A majority of the galaxies are UV weak, and  only $\sim$7\% of the galaxies in our sample have \fuvminusk bluer than 8.8 mag suggesting signs of recent star formation or bright accretion disks.  Furthermore, these FUV bright galaxies are not powerful in the radio ($P_{1.4GHz} < 10^{22}\,WHz^{-1}$) and are also less luminous in the \ks band and thus less massive.

We have examined the 13 FUV bright galaxies for evidence of star formation.  These fall into three categories:
	\begin{itemize}
		\item Galaxies with on-going star formation:  The two bluest galaxies (NGC~3413,  NGC~1705) are known to be undergoing a strong star burst (evidence from SDSS strong H$\alpha$ emission and \citep{annibali2003} respectively).  The radio power at 1.4 GHz for these galaxies is less than $10^{20}\,WHz^{-1}$ indicating that the FUV emission is dominated via star formation and not by AGN.   Also, NGC~855 shows CO emission \citep{nakanishi2007}, NGC~3928 has a starburst nucleus \citep{balzano1983} and IC5267 has a large number of star formation sites\citep{caldwell1991}, indicating ongoing star formation activity.  NGC~7252 is a merger remnant\citep{chien2010} that has old and new star forming population residing in the nuclear regions of the galaxy.

		\item AGN contribution:    NGC~5252 and NGC~5283 are AGNs with Seyfert type Sy1.9 and Sy2, respectively.  They show slight excess in the FUV light.  Similarly, NGC 4457 hosts a bright UV nucleus which is attributed to the central AGN \citep{flohic2006}.  

		\item  Unknown FUV origin:  In the rest of the galaxies, NGC~3955, NGC~4344,  NGC~4627,  and UGC~3097 do not have any strong evidence of ongoing star formation or AGN activity.  Thus far, the origin of the excess UV emission in these galaxies is unclear.

	\end{itemize}

\begin{figure}
	\centering
			\includegraphics[width=0.5\textwidth]{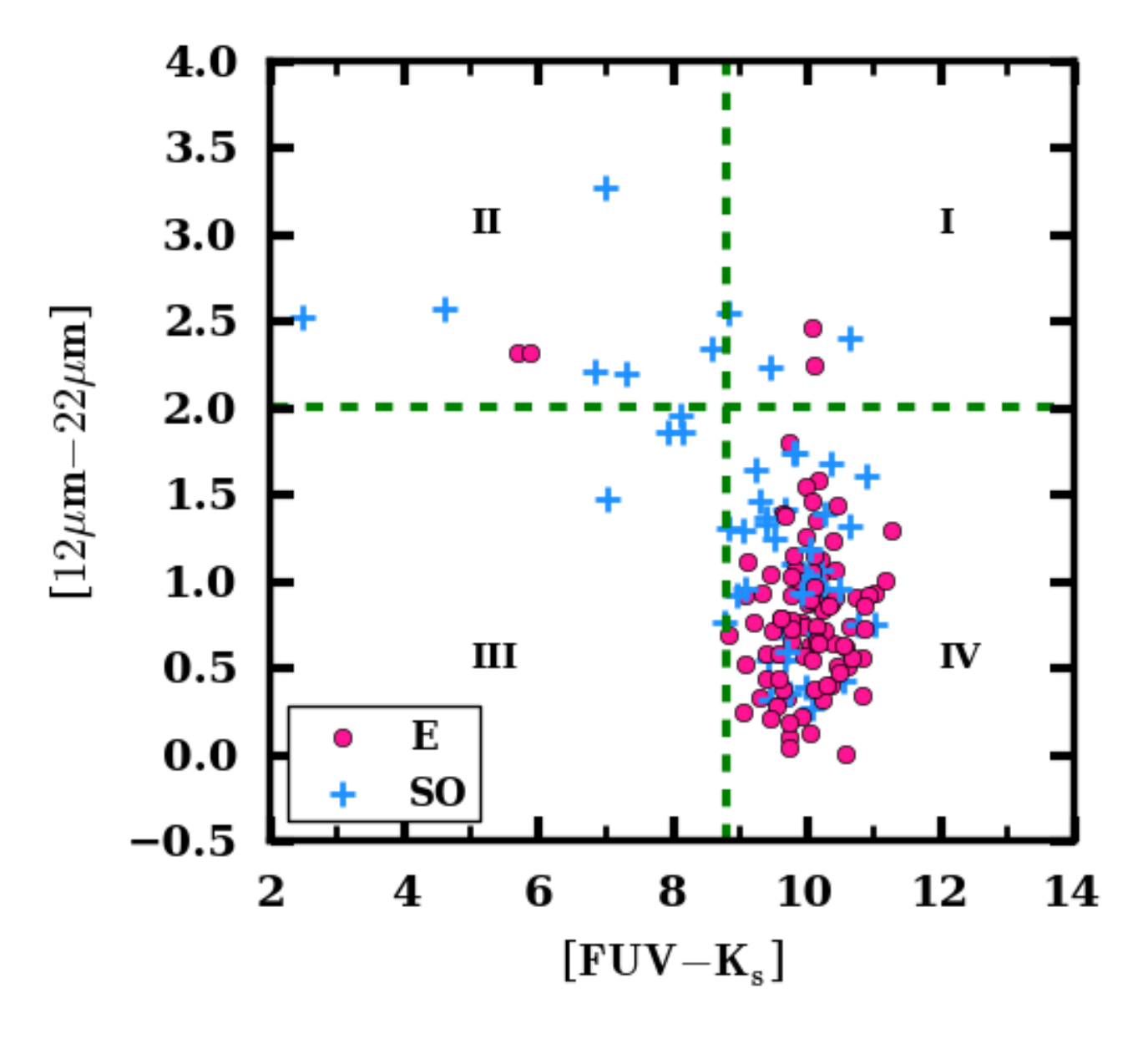}
			\caption{\textbf{Plot of \wise \wIIIminuswIV color versus \fuvminusk color}. The vertical green line is as defined in Figure ~\ref{fig:fuv-k_mk}.  The horizontal line defines  the redder galaxies.  Galaxies to the left of the vertical line (quadrant II and III) are star forming, those in the top right quadrant (quadrant I) are dust obscured star forming galaxies. Non-star forming galaxies tend to occupy the bottom right quadrant (IV).  Elliptical galaxies are colored in red and lenticular galaxies in blue. $\sim$7\% of the galaxies show indication of ongoing and obscured star formation.}
			\label{fig:w3-w4_fuv-k}
	\end{figure}

Similar to the UV, the mid-infrared (MIR) emission is also a good indicator of star formation in a galaxy.   A color-color diagram in  \wIIIminuswIV vs \fuvminusk for the galaxy sample is shown in Figure ~\ref{fig:w3-w4_fuv-k}.  Only 160 galaxies have the photometry for 12 $\mu$m, 22 $\mu$m, FUV and \ks band.   The green dashed vertical line at \fuvminusk = 8.8 mag  separates star forming and non-star forming galaxies.  We draw the horizontal line at \wIIIminuswIV = 2.0 mag to emphasize the concentration of galaxies centered at (10, 0.5).  The galaxies are color coded with ellipticals in pink circles and lenticular galaxies in blue plus symbols.  The four quadrants are named I, II, III and IV for convenience.  The galaxies in quadrant II and III ([\fuvminusk $<$ 8.8 mag)  are bright in the FUV and show signs of a young stellar population. The galaxies in quadrant I are bright in the IR but faint in the FUV, indicating star formation that is obscured by dust.  Galaxies in IV quadrant are redder in \fuvminusk and are not undergoing substantial star formation.  The majority of the galaxies that show star formation (i.e. in quadrants I, II and III) are lenticular galaxies.  \\

Using the  IR and FUV colors, we examine the star forming properties of our sample.  We notice a small fraction of star forming galaxies ($\sim$7\%) that are identified based on the FUV excess in the \fuvminusk and IR excess in the \wIIIminuswIV.  These star forming galaxies are also less massive and weaker in radio power than that of the galaxies without excess \fuvminusk.\\
\\


\subsection{SFR estimation using FUV}
		We now proceed to estimate the star formation rate (SFR) for our galaxy sample.  The most frequently used SFR indicators are UV continuum, recombination lines (primarily H$\alpha$, but H$\beta$, P$\alpha$, P$\beta$ have been used) , forbidden lines ([OII]$\lambda$3727), mid to far IR dust emission and radio continuum emission at 1.4 GHz   \citep{kennicutt1998}. The calibration of SFR for these different star formation tracers  are prone to systematic uncertainties from uncertainties in IMF, dust content and distribution and metallicity. However, the scaling relations offer a convenient method to compare the SFR properties in a large galaxy sample.   To estimate the SFR in our galaxy sample, we use the calibration in \cite{salim2007} which was derived to suit the GALEX wavebands.  This relation is valid in the $\lq$constant star formation approximation' where the SFR is assumed to remain constant over the life time of the UV emitting population ($<10^8$ year).  It also assumes a Salpeter IMF with mass limits from 0.1 to 100$M_{\odot}$.  \\
		
		\begin{figure*}
		\centering
			\includegraphics[width=0.9\textwidth]{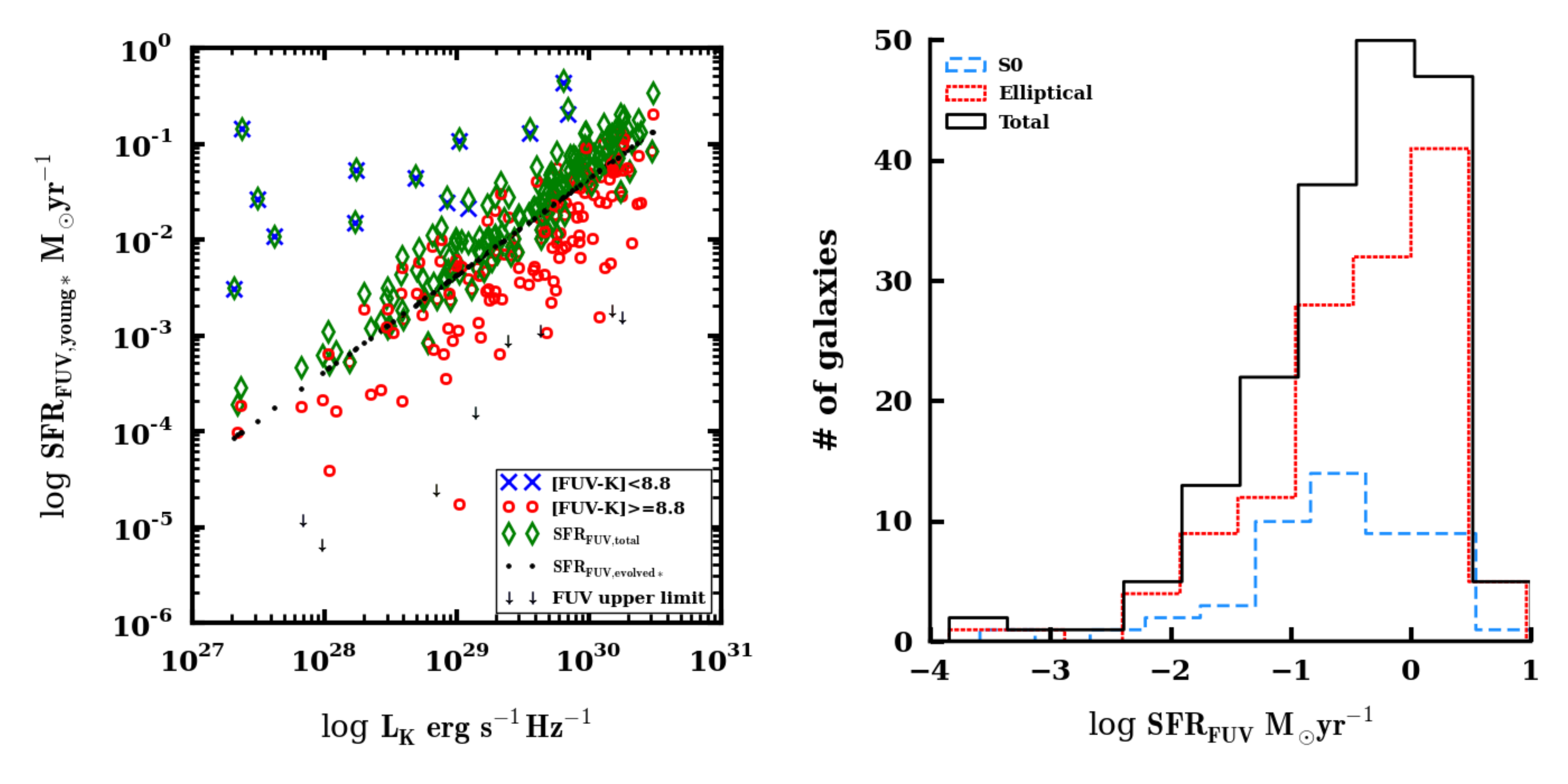}
			\caption{ \textbf{Left panel: Plot of SFR versus \ks band luminosity}.  The SFR is estimated using the FUV luminosity from which evolved stellar contribution has been removed and thus is a much better estimate of SFR from young stars.  These objects  are indicated as red circles and blue crosses depending on their \fuvminusk color.  The green diamonds shows the SFR estimated using the observed FUV without subtracting the evolved stellar component.  The black dots shows the derived SFR of evolved stars.  The down arrows indicate the SFR obtained using the FUV upper limit of  24.7 magnitude.
			 \textbf{Right panel: Histogram of the SFR}.  SFR for ellipticals and lenticular galaxies is shown in red and blue lines respectively and the total SFR is shown in black line.  The upper limits in the FUV are not considered in the making of the histogram.}
			\label{fig:sfr}

		\end{figure*}

		The FUV emission can be from young stars as well as the evolved stellar population and from the accretion disks of AGNs.  To account for the FUV luminosity ($L_{FUV}$) from the young stars alone, we use the following technique to remove the contribution from the evolved stellar population:  
We chose galaxies that have \fuvminusk above the median \fuvminusk and treat them as non-star forming galaxies (which is a fair assumption to make, since star forming galaxies are defined to occupy the region below \fuvminusk  $<$8.8 mag).  We then compare the \ks and the FUV luminosity.  A fit to the $L_{FUV}$ vs $L_{K_s}$ gives an indication of the amount of FUV emission from the evolved stars.  We obtained the following relation
		\begin{equation}
			\label{eq:fuv_evolved}
			\log L_{FUV,evolved*} = a\,\log L_{K _s}+ b,
		\end{equation}
		with a  =1.0072 and b = $-3.63$.
This fit is used to estimate the FUV contribution from the evolved stellar population for the rest of the galaxies and subtract it from the observed FUV luminosity.  This gives the FUV luminosity that is preferentially due to young stars ($L_{FUV,young*}$), which is then used to estimate the SFR.  For galaxies that have not been detected in the FUV, we use the \galex detection limit magnitude of 24.7.  FUV emission can also be contaminated by the AGN accretion disk especially from unobscured Sy1 galaxies.  There are five Sy1 in our sample and these have been removed in the SFR estimation.  Also, these Sy1 have \fuvminusk $<8.8$ mag.\\

	We estimate the SFR using the following relation from \cite{salim2007},
		\begin{equation}
		\label{eq:sfr}
			SFR (M_{\odot}yr^{-1}) = 1.08 \times 10^{-28} L_{FUV} \quad	(ergs^{-1}Hz^{-1}).
		\end{equation}
		The FUV luminosity, $L_{FUV}$ has been corrected for Galactic extinction alone (ignoring internal extinction due to dust).  Thus our SFR estimates can be considered as lower limits.  From here on, we label $SFR_{FUV,young*}$ as just $SFR_{FUV}$.  All the derived quantities (stellar mass, SFR, radio power and the absolute \ks band magnitude) are available online in machine readable format.  Table ~\ref{tab:derivedprop} lists these quantities for a subset of the sample.\\

		In Figure ~\ref{fig:sfr}, the left panel shows the plot of the SFR obtained using $L_{FUV,young*}$ against \ks band luminosity.  The SFR for our galaxy sample with FUV detections is less than 0.4 $M_{\odot}yr^{-1}$.  The green diamonds are the galaxies whose SFR is estimated using the total FUV luminosity, i.e., before subtracting the UV light expected from evolved stars.  Galaxies that are marked with blue cross are the star forming galaxies identified with \fuvminusk$<8.8$ mag.  There is an overlap of green diamonds and the blue crosses suggesting that there is no significant change in the SFR before and after subtracting the evolved stellar contribution.  All of the detected FUV emission is probably from young stellar population.  The red circles are the galaxies with \fuvminusk $\geq8.8$ and their SFR are less than the $SFR_{FUV,total}$ indicating a significant contribution to FUV from evolved stellar population.  The black dots show the SFR estimated from FUV using equation \eqref{eq:fuv_evolved}. The down arrows are the galaxies with FUV upper limits.  The right panel shows the distribution of $SFR_{FUV}$ for our sample.  The distribution is asymmetric and left skewed.  Most of the galaxies have SFR within 0.1-1 $M_{\odot}$/yr with a median at 0.4 $M_{\odot}$/yr and tails off at lower star formation rates.   The median SFR for ellipticals is higher at 0.5 whereas for the lenticulars, the median SFR is 0.2.  This probably is due to the fact that SFR correlates with the stellar mass \citep{brinchmann2004}, and the average (and the median) stellar mass for the ellipticals in our sample is greater than for the lenticulars.  \\

		\begin{figure}
		\centering
			\includegraphics[width=0.49\textwidth]{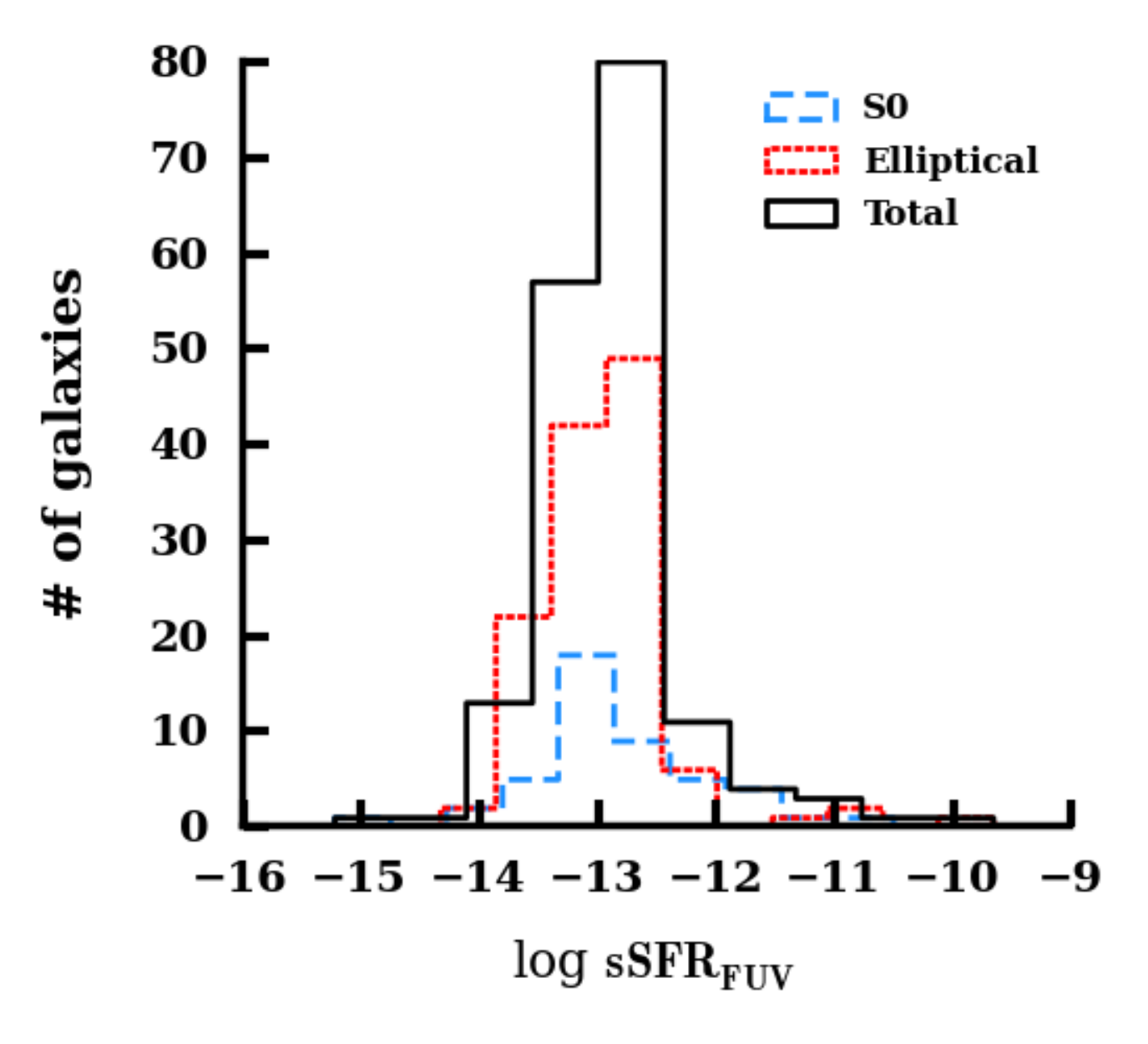}
			\caption{\textbf{Histogram of the specific SFR}.  sSFR for ellipticals and lenticular galaxies is shown in red and blue lines respectively. The total sSFR is shown with the black line.  The upper limits in the FUV are not considered in the making of the histogram.}
			\label{fig:ssfr}
		\end{figure}
		
		Figure ~\ref{fig:ssfr} shows the normalized distribution of the SFR per unit stellar mass, known as the specific SFR ($sSFR_{FUV}$).  It is color coded in red for ellipticals and in blue for lenticulars.  The median sSFR for ellipticals and lenticular is roughly  equal ($1.1\times 10^{-13}$ and $1.3 \times 10^{-13}$ respectively).  We perform a simple KS (Kolmogorov-Smirnov) statistical two-sample test to verify the claim (null hypothesis) that the distribution of the two population is the same.  The test gives a p-value of 0.05 and a KS statistic of 0.2.     The p-value gives us the probability of the strength of evidence against or in favor of the null hypothesis and the KS statistic tells the maximum distance between the cumulative distribution function of the two samples.  In this test, the small p-value suggests that there is only 5\% probability that the distribution of ellipticals and lenticulars appear to be same.  This suggests that there is significant difference between the distribution of the two populations.  


\subsection{Relation between radio power and host galaxy properties}
	\subsubsection{Radio power versus stellar mass}
	\label{Radio-Kband}
		  Galaxy luminosities in the \ks-band are 5 to 10 times less sensitive to dust than in the optical band, which allows them to be used as excellent tracers of stellar luminosity and thus stellar mass \citep{bell2001}.  Assuming  constant mass to light ratio, which is a good approximation to make especially in the \ks-band, the \ks-band luminosity gives an estimate of the stellar mass of the galaxy \citep[e.g.,][]{bell2003}.\\
  
\begin{figure}
	\centering
		\includegraphics[width=0.5\textwidth]{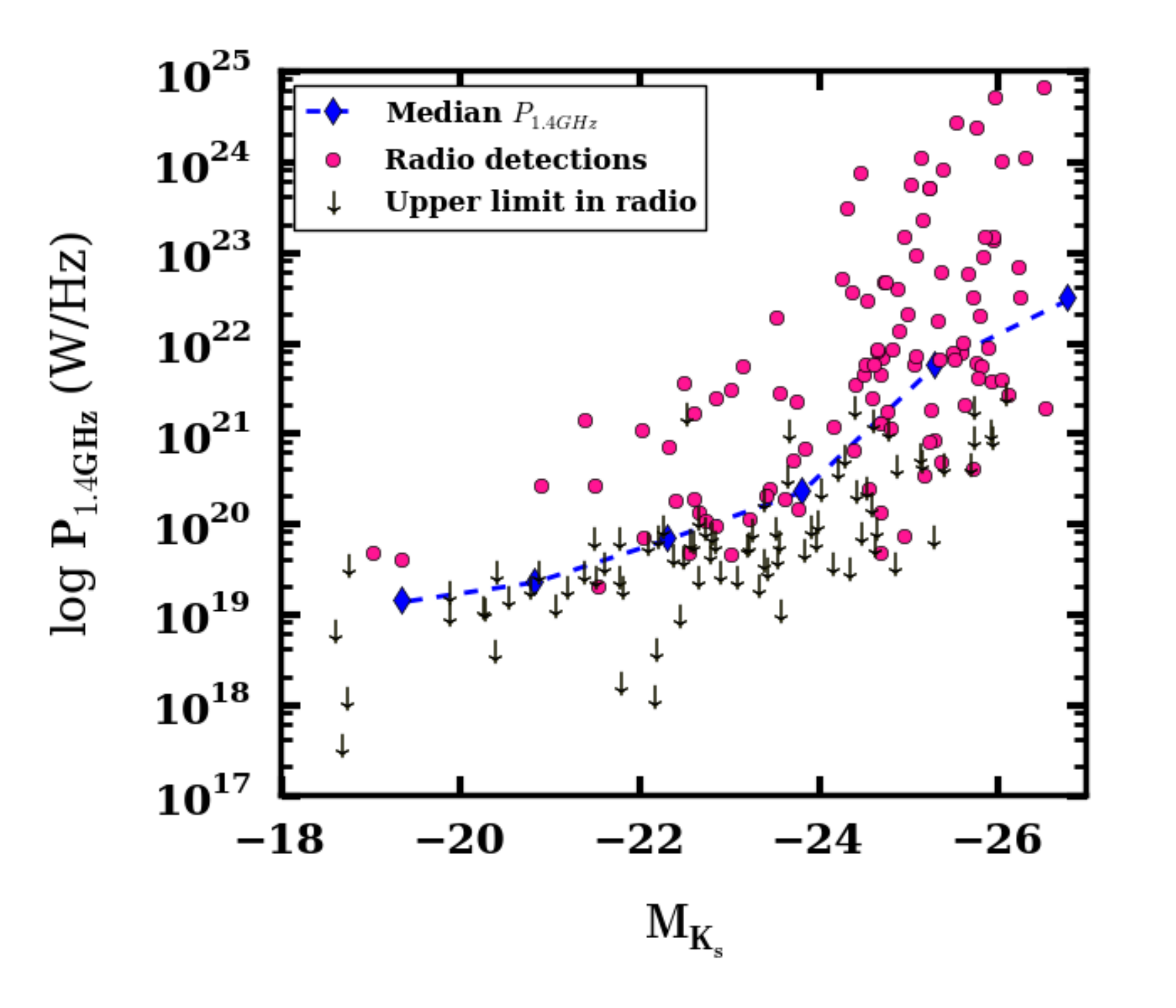}
		\caption{\textbf{Plot of total radio power (1.4 GHz) versus absolute k magnitude}.  Out of 231 sources, only 195 galaxies have \ks band magnitudes.  Sources with upper limits to the radio power are indicated with $\downarrow$.  About 56\% of the sources have radio flux measurements.  The dashed line shows the median radio power binned by the absolute \ks magnitude.  The median is calculated considering both the detected and undetected sources.   The plots shows that the upper envelope of radio power is a steep function of the total mass of the galaxy.  This indicates that massive galaxies are capable of hosting powerful radio sources.}
		\label{fig:radio_k}
	\end{figure}

  	Shown in Figure ~\ref{fig:radio_k} is a plot of 1.4 GHz radio power and absolute \ks band magnitude.   The radio power ranges between $10^{17}\,WHz^{-1}$  to $10^{25}\,WHz^{-1}$ and $M_{K_s}$ range from $-18$ to $-27$.  Galaxies that are fainter than 21 mag have radio power less than $10^{21}\, WHz^{-1}$.  Star forming galaxies are known to produce up to $10^{22}\,WHz^{-1}$ radio power at 5 GHz \citep{wrobel1991} which corresponds to $\sim$ $10^{21}\, WHz^{-1}$ at 1.4 GHz . The weakest radio source detected is NGC 855, and it has a radio power of $3.86\times10^{19}\, WHz^{-1}$ and $M_{K_s} =-19.4$.  This is a blue star forming dwarf elliptical galaxy \citep{nakanishi2007, walsh1990}.\\
    
\begin{figure*}
		\includegraphics[width=1.1\textwidth]{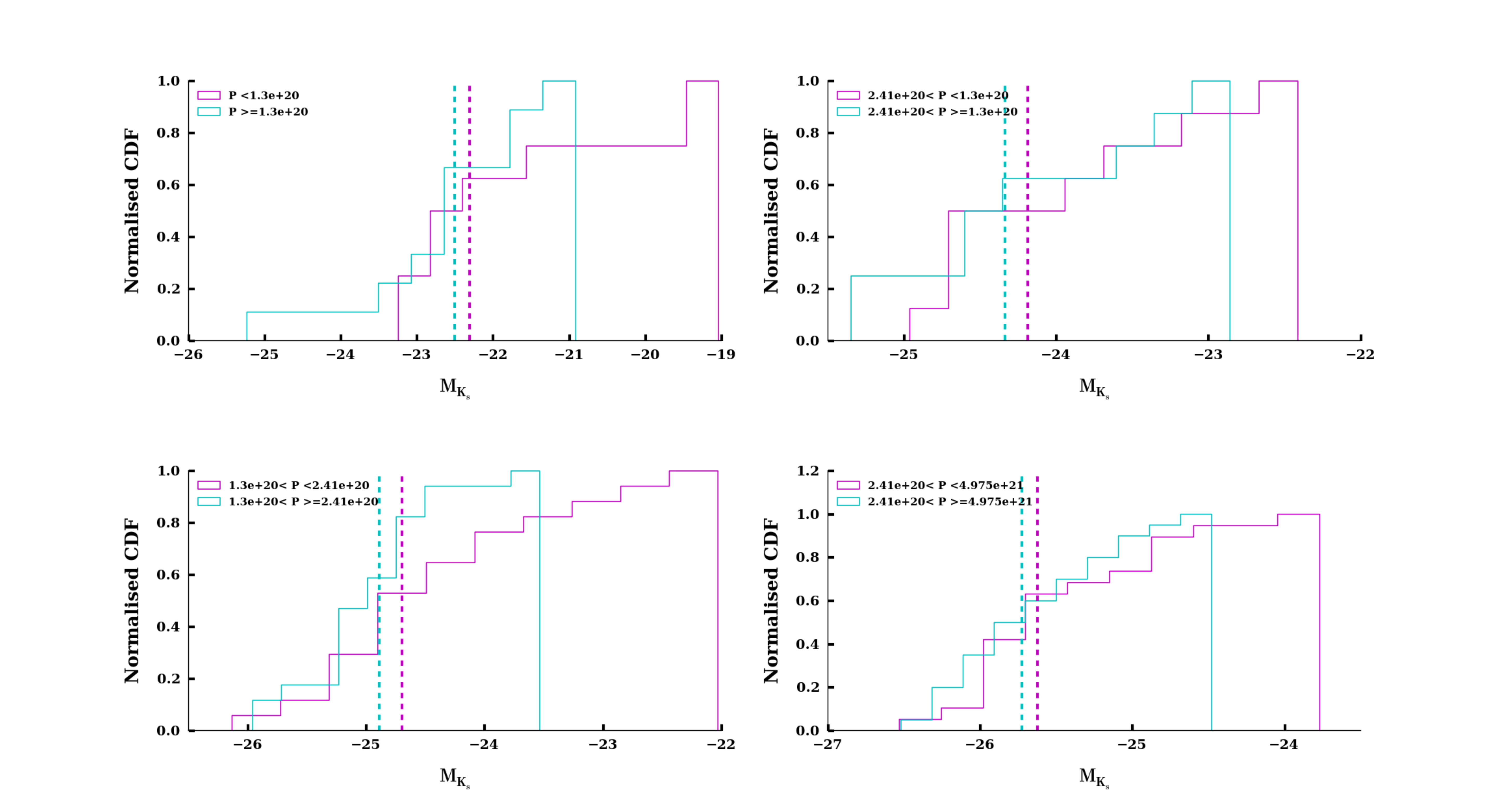}
		\caption{\textbf{Normalized cumulative distribution function of $M_{K_s}$ at different radio power bins and at different distance bins.}  The sample is divided into four equal distance bins.  Each panel corresponds to one of the distance bins and increasing from left to right.  Each panel shows the cummulative distribution function (CDF) of $M_{K_s}$ for galaxies with radio power greater than the median radio power in that bin (shown in cyan) and for galaxies with radio power less than the median radio power (shown in pink).  The cyan and the pink dashed lines represent the median $M_{K_s}$ for high radio bin and low radio bin respectively.}
		\label{fig:bias}	
	\end{figure*}

	Figure \ref{fig:radio_k} shows the relationship between the radio power and the stellar luminosity (mass).  Our results are consistent with previous investigations \citep{hummel1983, heckman1983, feretti1984, sadler1987, calvani1989, brown2011}.  But, a relation such as this between two luminosities need to be addressed carefully.   Since there is a tight correlation between luminosity and distance, any property that is related to distance can appear as a luminosity dependent relation (Malmquist Bias).  Before we claim to see any correlation between radio power and stellar luminosity, it is important to correct for this bias.  Following \cite{singal2014}, we examine whether the observed relation is due to Malmquist bias or is an intrinsic property of the sample.   We show in Figure \ref{fig:bias} the normalized cumulative distribution of $M_{K_s}$ at different radio power bins and at different distance bins.  We divided the sample into equal distance bins, except that we combined the last two bins into one bin due to the small number of sources.  In each distance bin, the sources are separated into two bins of radio power defined by the median radio power of the sample in that distance bin.  Each graph in the plot is the normalized cumulative distribution of $M_{K_s}$.  We notice that the median value of $M_{K_s}$, indicated by the dashed line, is higher for the higher radio power bin except in the third distance bin.  This suggests that the relation that we see in the Figure \ref{fig:radio_k}, i.e., luminous galaxies have higher radio power, is most likely intrinsic to the sample.\\

	 Compared to previous studies, our sample extends the investigation of radio power and galaxy absolute magnitude to fainter galaxies.  There is a broad distribution of radio power at a fixed \ks band absolute magnitude.  However, the two quantities show a strong correlation with a correlation coefficient of 0.75, and the probability of it arising by chance is $10^{-37}$.  There is an upper envelope of radio power that is a steep function of absolute \ks magnitude.  The median radio power (shown in the dotted blue line) also increases monotonically as a function of the galaxy brightness.  These results suggest that the maximum radio power from the galaxy is dependent on the mass of the galaxy.  Less massive galaxies appear to be capable of hosting only low radio power sources, while more massive galaxies are capable of hosting more powerful radio sources \citep{kauffmann2003, best2005, best2007}.  There is an apparent change in the slope of the median radio power around \ks magnitude of $-24$, which suggests that there may be two distinct processes that are responsible for the radio emission in a galaxy.  In the fainter galaxies, radio power can be attributed to star formation, i.e., a young stellar population going supernovae, whereas for massive galaxies, the radio power may be dominated by an AGN (see section \ref{radio_mir}).\\
	 
	This type of relation between galaxy mass and radio power is also found to exist in high redshift quasars \citep{browne1987, carballo1998, serjeant1998, willott1998, sanchez2003} and in radio galaxies \citep{yates1986, vanvelzen2014}.  Galaxies have to be massive enough to be powerful radio sources.  The fact that such a correlation exists for AGN and non-AGN population, from faint to bright galaxies is interesting.   Since the black hole mass scales with the bulge mass \citep[e.g.,][]{haring2004}, the correlation also suggests that the black hole mass closely relates to the maximum radio power \citep{franceschini1998, laor2000, liu2006}.  The plot also shows that there is a broad dispersion in radio power at a given absolute magnitude even for the most massive galaxies in our sample.   This is consistent with the hypothesis that high black hole mass is necessary but not sufficient to produce a powerful radio source.  Other physical parameters such as the spin of the black hole, accretion efficiency and other large-scale environmental effects may be responsible for the broad dispersion in radio power \citep[e.g.,][]{baum1995, meier1999, wold2007}.

	
		\subsubsection{Radio power and nuclear activity}
		\label{sec:radio_wise}
\begin{figure}
		\centering
			\includegraphics[width=0.5\textwidth]{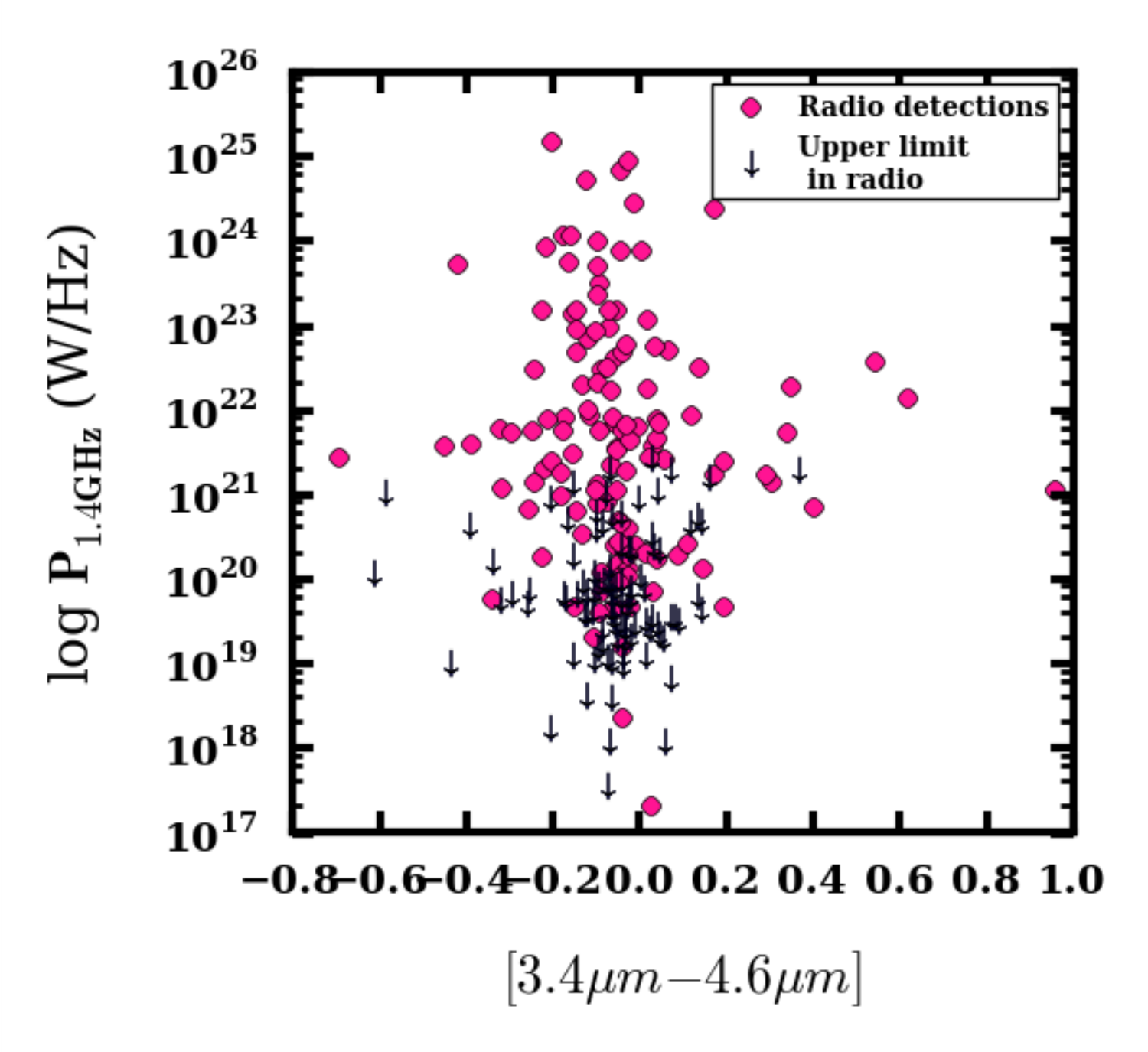}
			\caption{\textbf{Plot of radio power at 1.4 GHz versus WISE \wIminuswII color}.  Galaxies marked with down arrow have upper limit in the radio.  Most of the galaxies do not show a color excess.  This indicates that the galaxies in the sample are not associated with bright accretion disks.}
			\label{fig:radio_w1-w2}
		\end{figure}

			The origin of \wise mid-IR emission is associated with a combination of continuum emission from dust, atomic and molecular emission lines and features associated with PAHs that are heated by young stars and AGN, as well as Gyr old evolved stellar population \citep{jarrett2013}.  Shown in Figure \ref{fig:radio_w1-w2} is a plot of 1.4 GHz radio power against \wIminuswII infrared color.  The \wIminuswII color is sensitive to warm/hot dust and thus to optical/UV nuclear activity.  An excess in this color identifies galaxies in which hot dust surrounding the AGNs produces a strong mid-IR continuum that dominates the host galaxy emission \citep{stern2012}.  This enhanced mid-IR continuum may be  associated with the dusty torus heated by the radiation from an accretion disk.  About 83\% of the galaxies in our sample lie in a narrow color range (between $-0.3$ and 0.1 mag), and only a few galaxies show excess IR emission.  The fact that most of the galaxies do not show a color excess indicates that the galaxies in our sample are not associated with bright accretion disks, and if they are AGN, they are accreting in a radiatively inefficient process\citep[e.g. ADAF][]{narayan1994}.\\

	\subsubsection{Radio power and SFR relation}
\begin{figure}
			\centering
			\includegraphics[width=0.5\textwidth]{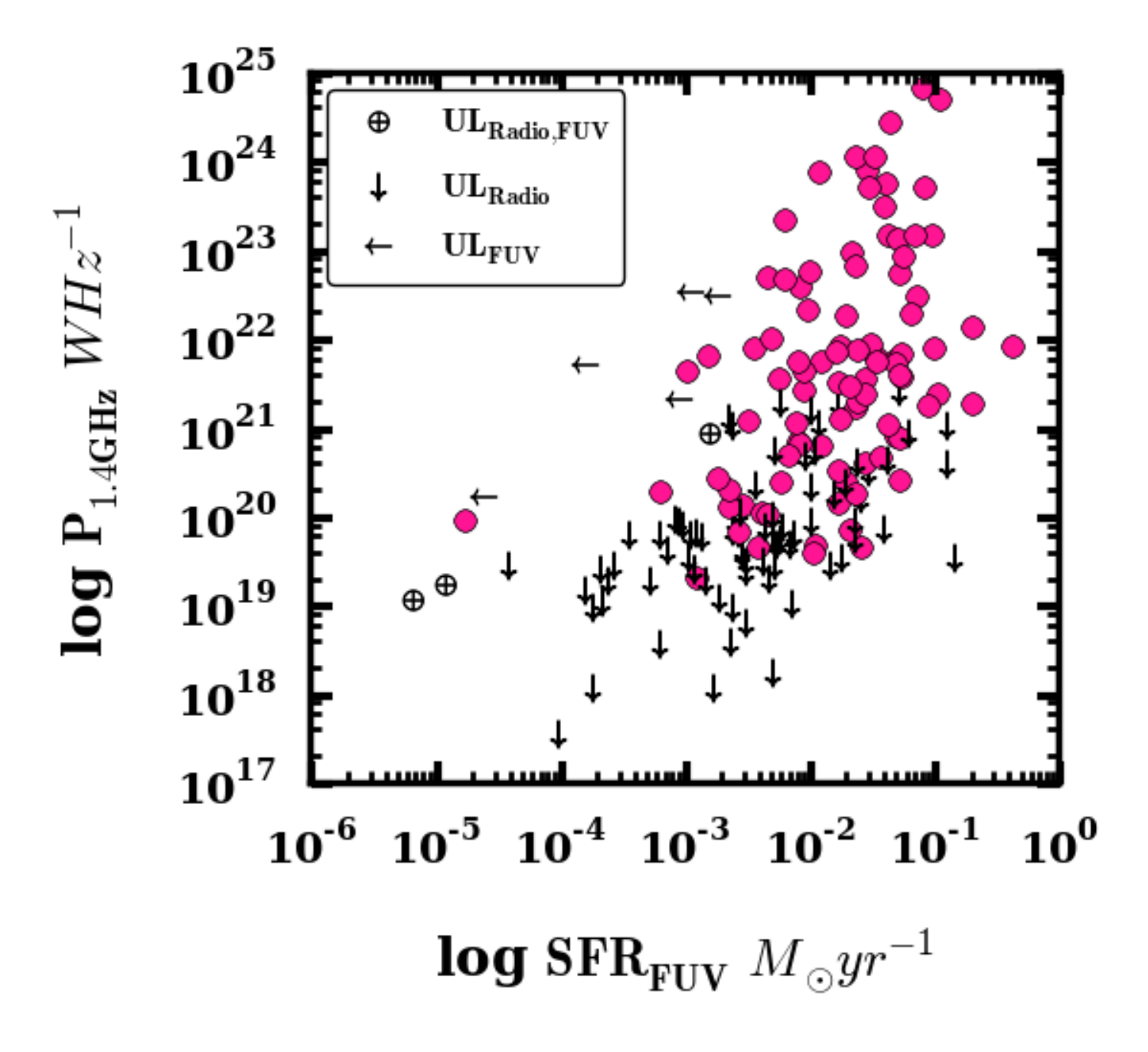}
			\caption{\textbf{Relation between radio power  and the estimated rate of star formation.}Upper limits in the radio are shown with a down arrow, FVU with a left arrow and upper limits in both FUV and radio are shown with an oplus symbol.   Radio detections are shown  in pink.  The observed weak correlation between the radio power and SFR is likely due to the correction of both radio power and SFR with galaxy mass (Figure ~\ref{fig:radio_k} and Figure ~\ref{fig:sfr_mass} respectively).}
			\label{fig:radio_sfr}
		\end{figure}

		 The supply of cold gas to fuel star formation and AGN activity can include  galaxy mergers/interactions,  the cooling of gas in the ISM or surrounding hot halo, and the mass loss from stars \citep[e.g.,][]{heckman2014}.    The relation between radio power and SFR can provide some insight into the source of the fueling (Figure ~\ref{fig:radio_sfr}).   \\

\begin{figure}
		\centering
			\includegraphics[width=0.5\textwidth]{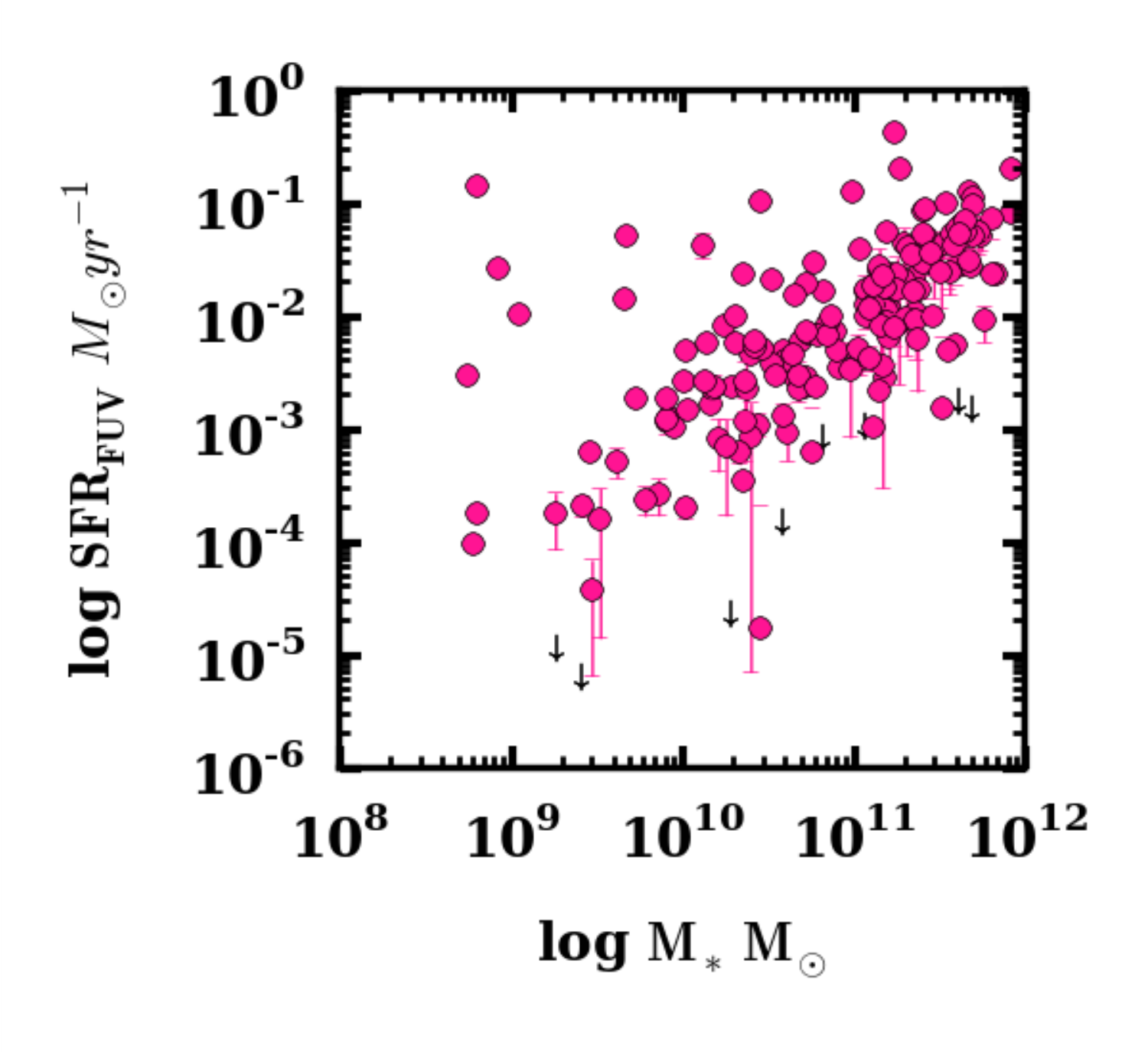}
			\caption{\textbf{Plot of SFR against stellar mass of our sample.}  }
			\label{fig:sfr_mass}
		\end{figure}

	We see a clear correlation (spearman correlation coefficient of 0.45 at a significance level of $10^{-6}$)  although with a bit of scatter, between radio power and the SFR (Figure~\ref{fig:radio_sfr}).  Figure ~\ref{fig:radio_k} shows a weak correlation between radio power and galaxy mass, while Figure ~\ref{fig:sfr_mass} shows a correlation between SFR and galaxy mass \citep[see also][]{brinchmann2004}. This suggests that the weak correlation between radio power and SFR may be  due to a correlation of both radio power and SFR with galaxy mass.  Such a correlation with galaxy mass would be consistent with an origin of the gas supply which fuels the AGN and star formation associated with the host galaxy itself rather than a predominantly external origin (e.g., major mergers). In this case the gas supply might be due to stellar mass loss \citep[e.g.,][]{faber1976, knapp1992, voit2011} or perhaps cooling from the ISM or halo \citep{binney1981, forman1985, canizares1987, voit2015}.\\
	
	The figure also shows a broad dispersion of several orders of magnitude between the radio power and SFR (especially between 0.01 and 0.1 $M_{\odot}yr^{-1}$).  A similar large dispersion is observed in the radio power vs stellar mass relation (Figure \ref{fig:radio_k}), which indicates that, producing a radio source is a complicated process that depends not only on the gas supply but also on the gas transport mechanism, black hole spin and black hole mass, accretion rate and the external environment \citep[e.g.,][]{baum1995, meier1999, wold2007}.  These will naturally add dispersion to the relation between radio power and the SFR.  In addition irregular fuel supply \citep{tadhunter2011, kaviraj2014} and variability in the AGN $\lq$on' phase \citep{hickox2014} will further weaken the correlation. \\
	
	Determining the connection between AGN feedback and star formation from this relation is not straightforward.  Although we notice that galaxies that have low SFR do not have high radio power, it is not clear whether AGN is responsible for the low SFRs by suppressing the star formation via feedback.  If AGN feedback is ongoing and has suppressed the star formation even though the radio power is low, this would imply that radio power is not a good indicator of AGN feedback.  There can be other possibilities for the observed low radio power at low SFRs.  Absence of major mergers (as suggested above) can leave a galaxy with less cold gas, that is insufficient for high SFR and power nuclear accretion. 	
	
	\subsubsection{Radio-MIR flux correlation}
		\label{radio_mir}
		\begin{figure*}
		\centering
		\includegraphics[scale=0.5]{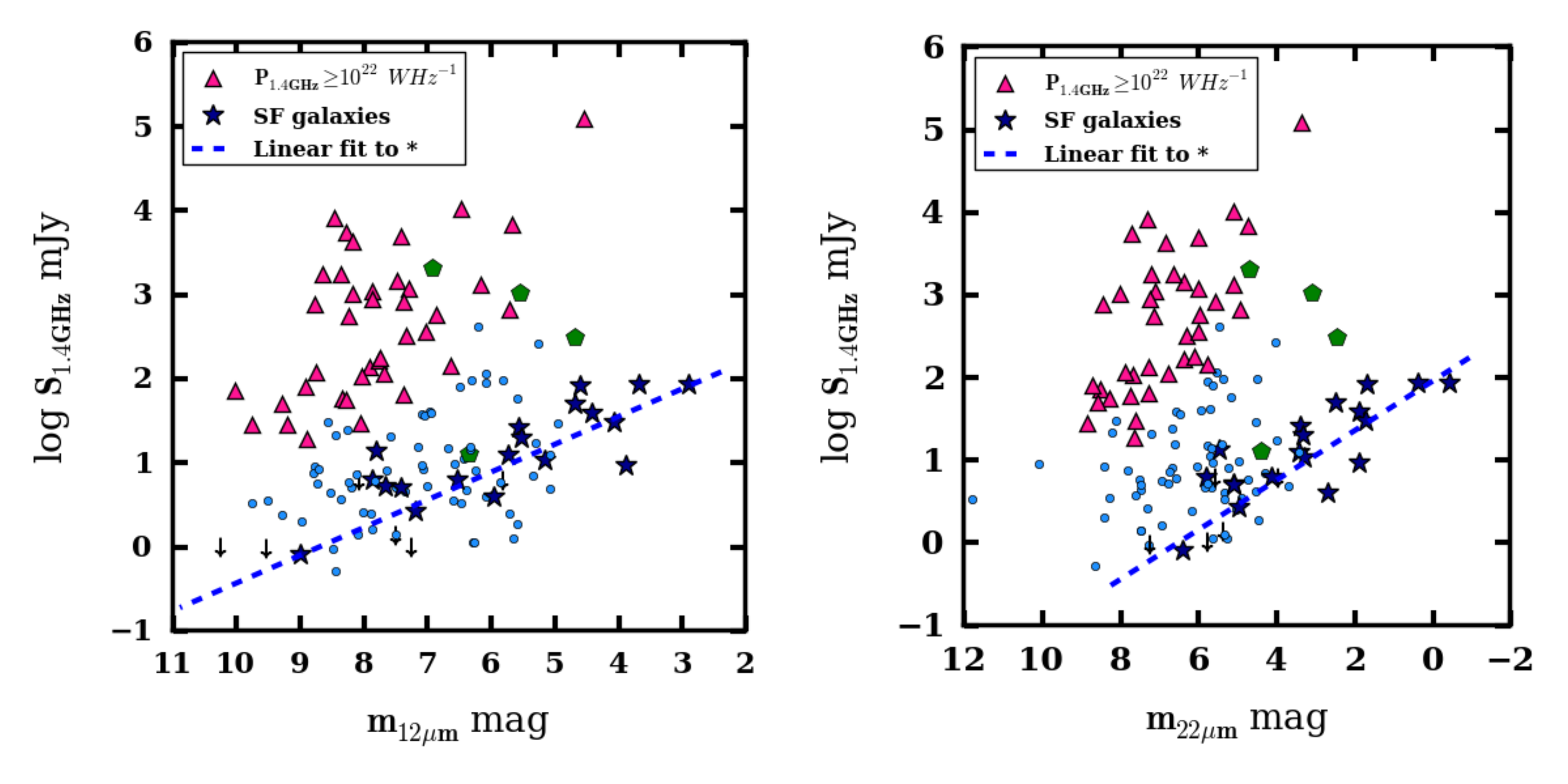}
		\caption{\textbf{Radio flux versus \wise mid-IR apparent magnitude}.  
The blue * are those galaxies that are identified as star forming using the criterion $[FUV-K] < 8.8$ and $[W3-W4] > 2.0$.  The pink triangles are galaxies that have radio power greater than $10^{22}\,WHz^{-1}$ and clearly fall off the radio-MIR relation determined by the linear fit.   The green pentagon symbols represent galaxies that are both star forming and have $P_{1.4GHz}\geq10^{22}\,WHz^{-1}$.  The rest of the galaxies are indicated with pale blue dots.  The blue dashed line is the least square regression fit to the star forming galaxies.   We have excluded the green points in the fitting process as their behavior deviates from the blue points.  The down arrows are star forming galaxies with an upper limit in the radio.}
		\label{fig:radio_w3flux}
	\end{figure*}
	
		Radio emission observed in a galaxy is non-thermal from relativistic electrons accelerated either by the AGN or by the supernovae \citep{dejong1985, condon1992}.  One way to identify the origin of the radio emission is to compare the radio flux with the far-IR flux.   A correlation between radio and infrared emission suggests that sources that are responsible for IR emission, are also responsible for emission in the radio.  Young stars with $M\sim8M_{\odot}$ and above emit most of their energy in the UV which is then absorbed and re-radiated in the IR by the dust.  At the end of their life, these massive stars explode as supernovae which accelerate the electrons to relativistic speeds resulting in radio emission due to synchrotron.  Thus, the relationship between radio and IR emission can trace star formation activity \citep{dejong1985, helou1985}.\\

	Figure ~\ref{fig:radio_w3flux} shows the relationship between 1.4 GHz radio flux and \wise mid-IR apparent magnitudes.    At both  12 and 22 \micronm, we find a correlation similar to the radio-FIR (see also
\citet{appleton2004}).   We compute the radio-MIR regression coefficients using Kaplan-Meier method so as to consider the upper limits in the radio during the fit.  We used iraf task \textit{buckleyjames} for this purpose and obtained the following relation:\\
	\begin{align}
		\log S_{1.4GHz} &= -0.33(\pm0.04) m_{12\mu m} + 2.87 \\
		\log S_{1.4GHz} &= -0.30(\pm0.04) m_{22\mu m} + 1.96,
	\end{align}
	where $S_{1.4GHz}$ is the radio flux at 1.4 GHz and $m_{12\mu m}$, $m_{22\mu m}$ are the apparent magnitudes at 12 and 22$\mu$m waveband respectively.  The correlation coefficient for both the relations is 0.83 and the dispersion in the regression is $\sim$ 0.35 dex.  In terms of the MIR flux, the slope of the radio-12 \micronm and radio-22 \micronm relation is $\sim$ 0.8 and 0.75 respectively which are comparable to the slope obtained in the previous studies of radio-MIR \citep{gruppioni2003} and radio-FIR relation (which is $\sim$ 0.9).  Unlike the tight radio-FIR correlation, the radio-MIR relation has a high dispersion \citep{appleton2004}, possibly because the $12\mu m$ emission is not solely due to the dust heated by young stars, but  can arise from the PAHs heated by young/evolved stars/AGN or from the dust shells of AGB population.\\

	  Galaxies shown in pink triangles in Figure ~\ref{fig:radio_w3flux} are the galaxies with $P_{1.4GHz}\geq 10^{22} \, WHz^{-1}$.  Although this radio power cut was chosen arbitrary (but with the knowledge that star forming galaxies have $P_{1.4GHz}\sim 10^{21} \, WHz^{-1}$ \citep{wrobel1991}),  these galaxies show a noticeable departure from the normal radio-MIR and follow a different relation with a slope of 0.4.  Despite the huge scatter,  we notice a trend in the radio-MIR relation.  These galaxies show an excess radio emission relative to its MIR flux, which can be attributed to AGN origin.  Thus, these galaxies are potential candidates for being an AGN.  We define $10^{22}\,WHz^{-1}$ as the threshold radio power and for the rest of the sections, we define galaxies above this threshold as \lq radio bright' and below the threshold as \lq radio faint' galaxies.  \\
	  
	  The rest of the galaxies are identified by pale blue dots.  These are fainter than 5 mag in 12\micronm and 4 mag in 22\micronm.  Although with a large scatter, these galaxies appear to follow the radio-12\micronm  relation.  On the other hand, in the radio-22\micronm plot, these galaxies do not seem to follow the radio-22\micronm relation for star forming galaxies.  Also, these fall in the region that is mid-way between the star forming galaxies and the radio bright galaxies.  About 28\% of the radio faint galaxies are star forming galaxies and fall on the radio-MIR correlation. \\


	\subsubsection{Specific SFR and stellar mass}
\begin{figure}
		\centering
			\includegraphics[width=0.5\textwidth]{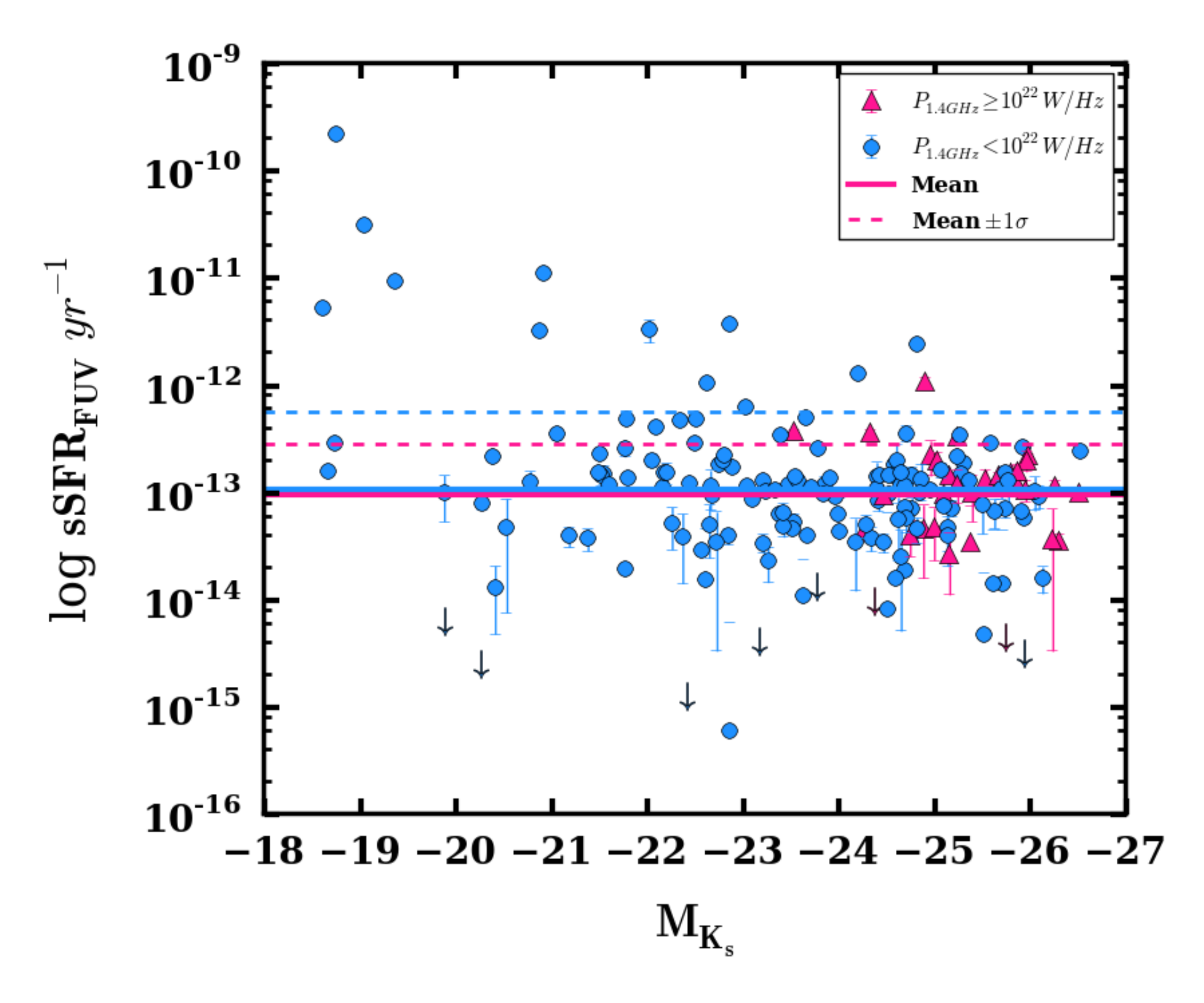}
			\caption{\textbf{Specific SFR vs absolute \ks band magnitude}.  Pink triangles indicate galaxies that have an excess radio power ($P_{1.4GHz} \geq 10^{22}WHz^{-1}$) and  the blue circles indicate low radio power ($P_{1.4GHz} \le 10^{22}WHz^{-1}$) galaxies.  The average sSFR is indicated with solid line and the deviation from the mean is indicated with dashed line.  The average sSFR for both the groups is small and almost equal.  The results indicate that galaxies are not experiencing significant growth or significant AGN feedback.}		
			\label{fig:ssfr_k}
		\end{figure}

		Specific star formation rate (sSFR), which is the SFR normalized by the stellar mass, traces the star formation efficiency.  The sSFR indicates fractional galaxy growth due to star formation.   Figure ~\ref{fig:ssfr_k} shows the sSFR with respect to the absolute \ks magnitude.  Radio bright galaxies with $P_{1.4GHz}\geq10^{22}\,WHz^{-1}$ are indicated with pink triangles.  The mean and one sigma deviation above mean for the sSFR is shown in solid and dashed lines respectively.  The mean sSFR for the radio bright and radio faint galaxies is nearly the same.  Most of the galaxies have a low sSFR of $10^{-13}\,yr^{-1}$.  The least square regression for the two quantities gives a flat slope slope of $10^{-12}$.  This flat relation suggests that the sSFR is not a strong function of  galaxy stellar mass. \\

There are a few outlier galaxies (above the dashed lines) that show increased sSFR which is observed in both galaxy populations i.e. radio bright and radio faint galaxies.  The standard deviation in the sSFR is noticeably different in the two populations.  But the proportion of high to low sSFR (i.e. above and below the dashed line) in radio bright galaxy is nearly the same as that in radio faint galaxies.  A statistical test using z-score proportionality is calculated for these two populations on the null hypothesis that the two population proportions are the same.  The test statistic gives a p-value of 0.4 indicating that the null hypothesis cannot be rejected.  What this tells us is that, the fraction of star forming galaxies in high radio power galaxies is similar to the fraction of galaxies forming stars in low radio power galaxies.  In addition, the KS two-sample test gives a high p-value of 0.9 indicating that the two samples are drawn from the same distribution further supporting the idea that the distribution of the sSFR in both the samples is nearly the same. Thus, we find that star formation efficiency is small and independent of radio power in our sample. This suggests that our sample galaxies are not experiencing significant growth or significant AGN feedback.  \\\\
		
	\subsection{Comparision to galaxy clusters}
	Observational evidence exists for AGN feedback in action in the Brightest Cluster Galaxies (BCG) in the form of X-ray cavities \citep{mcnamara2007, fabian2012} and shocks \citep{mcnamara2005, fabian2006, forman2007} in the ICM .  Here we compare our sample with the Brightest Cluster Galaxies (BCG) in cool cores.		
									   We selected the BCGs from \citep{odea2008} (hereafter Odea08) and used their 1.4GHz radio data and IR derived SFR.  The Odea08 sample consists of BCGs that are located in the cores of X-ray luminous clusters that have optical line emission, thus preferentially selecting BCGs in cool cores. We also considered BCGs from \citep{rafferty2006} (hereafter R06) sample.  The R06 sample consists of BCGs that show evidence of X-ray cavities indicating AGN feedback.  The R06 data set provides  X-ray cavity power and mass cooling rate which are related to AGN jet power and the rate of star formation respectively.  The cavity power scales to jet radio power according to the following relation (Equation 1 of \citep{cavagnolo2010} ) :
			\begin{equation}
				\log P_{cav} = 0.75(\pm 0.14) \log P_{1.4} + 1.91(\pm 0.18),
			\end{equation} 
			and the average mass cooling rate is $\sim$ 4 times the SFR (section 4.3 of \citep{rafferty2006}) which is within the range of 3-10 suggested by \citep{odea2008}.\\

\begin{figure}
	\centering
		\includegraphics[width=0.5\textwidth]{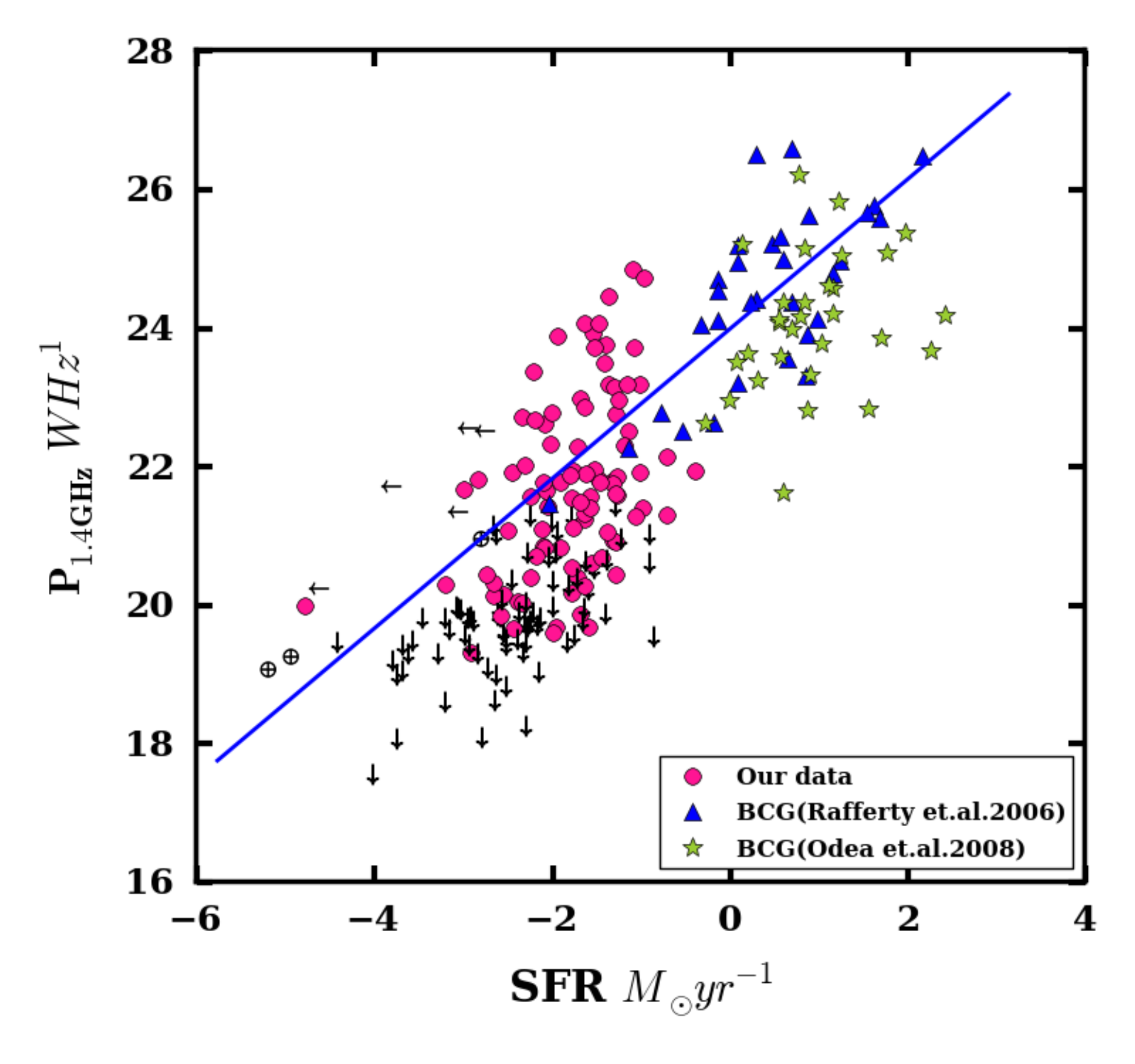}
		\caption{\textbf{Radio power at 1.4GHz versus SFR - Comparing our sample with the BCGs}.  The markers in pink and black are the galaxies in this study where detections are indicated in pink and upper limits with black arrow.  The blue triangles and green stars are the BCGs from \citep{rafferty2006} and \citep{odea2008}.  The solid blue line is the best fit line ($\log P_{1.4GHz} = 1.08\log SFR+24.0$)to the BCGs in R06 sample. The mean radio power of the BCGs is $\sim10^{24}\, WHz^{-1}$.  The spread in the correlation suggests various possibilities such as different sources of gas supply,  black hole spin, accretion rate and the time delay between the triggering of star formation and AGN activity.}
		\label{fig:radio_sfr_bcg}
	\end{figure}

	  In Figure ~\ref{fig:radio_sfr_bcg} we plot the radio power as a function of SFR extending the relation to the BCGs.   Our galaxy sample follows broadly along the R06 line thus extending the relation from weak radio power galaxies to more radio powerful galaxies in clusters .  It covers about eight orders of magnitude in both radio power and in SFR. However, there is a scatter about this relation which spans roughly four orders of magnitude.  A Spearman test for the combined sample indicates a strong correlation with a correlation coefficient of 0.78 and the probability of it arising by chance is $\sim 10^{-32}$.  Thus, the plot shows a general trend between SFR and the radio power across several orders of magnitude.  The relation in the plot suggests that the fuel supply for the  triggering of star formation and the AGN has a common origin.  We suggest that the same factors that introduce scatter to the  radio power vs stellar mass relation, contribute to the scatter here as well, which include gas supply mechanisms, the amount of gas available for fueling AGN and star formation, and the relative differences in life cycles of the star formation and AGN activity.   Although these factors contribute to the scatter, the relative contribution of these factors remains uncertain.\\
	
	The average radio power of the BCGs is  $\sim 10^{24}\,WHz^{-1}$.  For an effective mechanical feedback via heating, the radio source should have $P_{1.4GHz}\sim10^{24}-10^{25}\,WHz^{-1}$ \citep{best2006}.  The majority of the galaxies in our sample have radio powers below this value.  These results are consistent with weak (or negligible) AGN feedback in our sample. 
	
	\subsubsection{Comparision of sSFR}	
	To estimate the total stellar mass for the Odea08 sample, we used the flux measurements at IRAC bands 3.6 \micronm and 4.5 \micronm.  The redshift range of this sample is between 0.017 to 0.25.  The redshifted light at the these central wavelengths fall within the IRAC bandwidth. A simple and yet robust conversion between the stellar mass and infrared flux is given by \citep{eskew2012}:
	\begin{align}
		M_* = 10^{5.65} F_{3.6}^{2.85}F_{4.5}^{-1.85}\left(\frac{D}{0.05}\right)^2
	\end{align}
	
	where $M_*$ is  in $M_{\odot}$, D is in Mpc, $F_{3.6}$ and $F_{4.5}$ are in Jy.
We used the luminosity distance estimates to these BCGs from NED with cosmological parameters for $H_o$=71 $km s^{-1}Mpc^{-1}$, $\omega_{matter}$=0.27 and $\omega_{vaccum}$=0.73.  The $F_{3.6}$ and F$_{4.5}$ estimates are taken from \citep{quillen2008}.  \\

\begin{figure}
	\centering
		\includegraphics[width=0.5\textwidth]{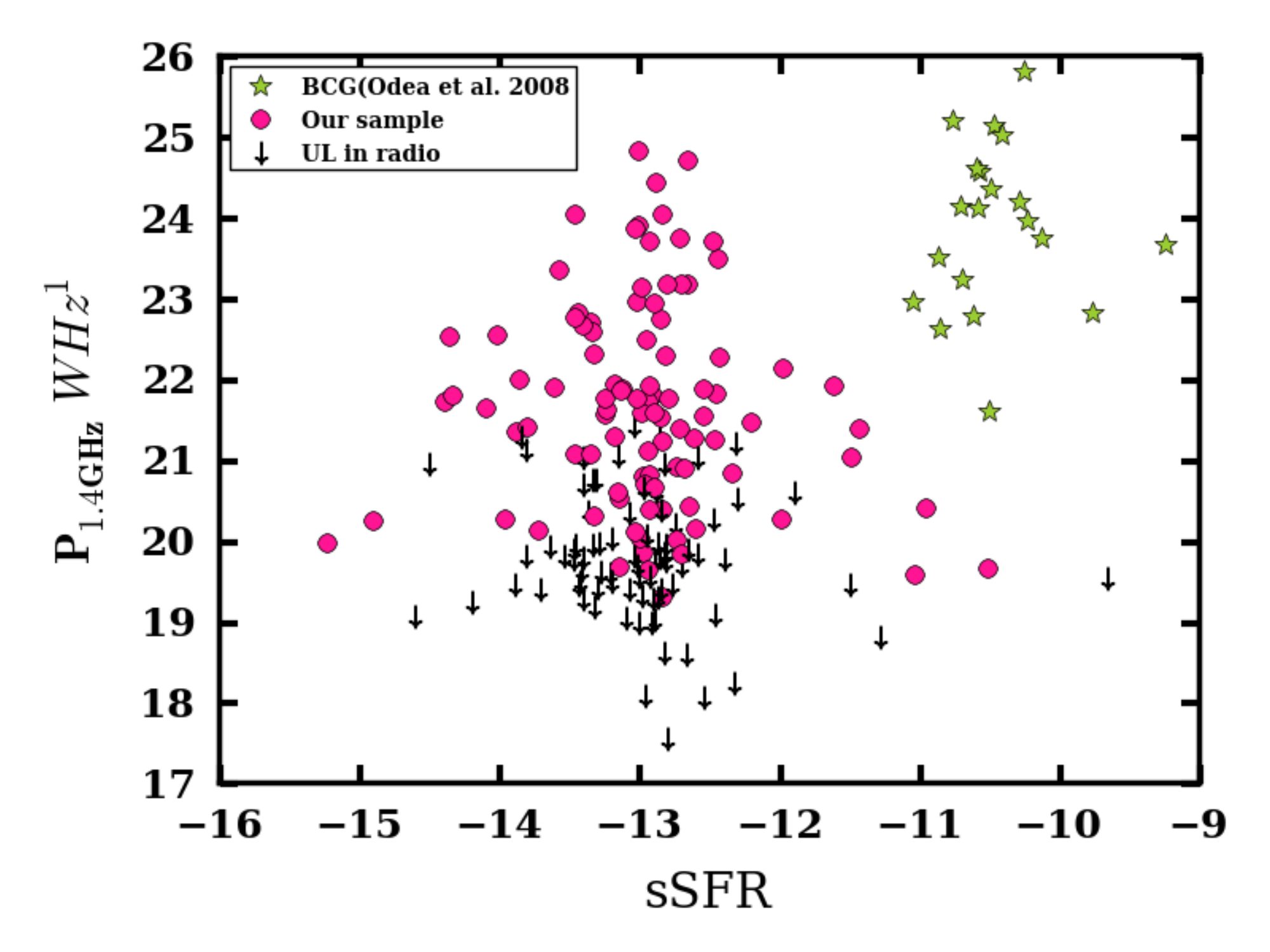}
		\caption{\textbf{Plot of radio power versus sSFR - Comparing our sample with Odea08 sample}.  The description of the legend is same as that of Figure \ref{fig:radio_sfr_bcg}.  }
		\label{fig:bcg_radio_ssfr}
	\end{figure}

The average stellar mass we obtained for Odea08 sample is $2.3\times10^{11}\,M_{\odot}$, which is comparable to the average stellar mass of BCG at low redshifts \citep{liu2012, fraser2014}.    We plot the radio power against the sSFR  in Figure \ref{fig:bcg_radio_ssfr}.  The average sSFR for the BCG is $3.6\times10^{-11}\, yr^{-1}$ which is roughly two orders of magnitude higher relative to the local early type galaxies. We see that both the star formation efficiency and the radio power are higher in the BCGs than in our sample. This suggests that although feedback is likely present in the BCGs \citep[e.g.,][]{fabian2012}, it is not sufficient to completely suppress star formation \citep[e.g.,][]{odea2008, tremblay2012b, tremblay2015}.

\section{Summary}
\label{summary}
	We collected multiple wavelength data (radio, IR and UV ) for a sample of 231 early type galaxies at $z<0.04$ and analyzed the properties of star formation and radio mode feedback.   The main results are as follows. 
\subsection{Properties of Star Formation}

The SFR in our sample tend to be low ($< 1M_{\odot}yr^{-1}$) and only 7\% of the galaxies show obvious signs of ongoing star formation via the  \fuvminusk and \wIIIminuswIV colors. In addition, the sSFR is very small.  The star forming galaxies trace a radio-MIR correlation similar to that seen in other samples of star forming galaxies.  These results indicate that galaxy building in early type galaxies has essentially ceased at the present epoch.
		
\subsection{AGN Radio Properties }

The Radio-MIR relation shows that galaxies with $P\geq10^{22}\,WHz^{-1}$ have radio power in excess of that expected to be produced by the estimated star formation rates and thus are potential candidates for being radio AGN.  Only $\sim 20\%$ of the galaxies in our sample have $P\geq10^{22}\,WHz^{-1}$. Only a few of the high radio power galaxies  show excess 4.6 \micronm flux, an indication of hot dust heated by an accretion disk.  This indicates that the majority of the radio AGN are accreting gas in a radiatively inefficient manner \citep{ho2009}.  \\

 There is an upper envelope of radio power that is a function of galaxy stellar mass (and thus BH mass) suggesting that the maximum radio power scales with galaxy (BH) mass.  The large scatter in the relation between radio power and galaxy stellar mass suggests that high black hole mass is  necessary but not sufficient for producing a radio loud AGN. This is consistent with additional parameters (such as BH spin, accretion rate) playing an important role in determining radio power.
		
	\subsection{Relation between Radio and Star Formation Properties}
		
The sSFR is roughly independent of radio power in our sample, suggesting that radio mode feedback is not having a significant effect on star formation efficiency in these galaxies.  Alternately, radio power may not be a good proxy for radio mode feedback or the feedback is episodic.\\

The correlation between radio power and SFR is weak, and if real may be due to a correlation of both radio power and SFR with galaxy stellar mass.  This would suggest that the host galaxy is the source for the fuel (e.g., stellar mass loss for lower mass galaxies and cooling from the ISM/halo for more massive galaxies) for star formation and AGN activity in these galaxies.\\

 Two samples of cool core BCGs lie on the same relation for radio power and SFR as our sample over a range of eight orders of magnitude.  Although both star formation and radio mode feedback are constrained to be very low in our sample, the BCG samples exhibit both at high levels.  The relatively low radio power in our sample compared to the average radio power of the BCGs (i.e. $\sim10^{24}\, WHz^{-1}$) suggest that there may be a threshold in the radio power that is needed for the feedback from the AGN to affect the star formation in the host galaxy.


\section{Acknowledgements}
	This work was supported in part by the Radcliffe Institute for Advanced Study at Harvard University.  We thank Dr. Christine Jones for constructing the sample and sharing it with us.  We thank Dr. Bill Forman for his invaluable comments and suggestions which helped us to improve the paper. We are grateful to the facilities of Harvard-Smithsonian Center for Astrophysics (CfA), Cambridge where the project took its initial shape.  This work made use of the archival data from \galex, \twomass, \wise, NRAO and SUMSS and supplemented with information from Hyperleda and NASA/IPAC Extragalactic Database (NED).  The National Radio Astronomy Observatory (NRAO) is operated by Associated Universities, Inc., under cooperative agreement with the National Science Foundation.  We thank the NSF funded REU (Research Experience for Undergraduate) program in the Chester F. Carlson Center for Imaging Science at RIT. We also wish to acknowledge helpful conversations with colleagues. 
				
\clearpage				
\newpage


\begin{table}
\caption{Observational properties of a subset of the sample.}
\label{tab:sample}

\medskip
\resizebox{\dimexpr \linewidth+10cm\relax}{!}{%
\tabcolsep=2pt

\begin{tabular}{lccccccccccccccc}
	\toprule
		Name & RA & DEC & z & Vel & T & l\_F1.4 & F1.4 & W1 & W2 & W3 & W4 & $K_s$ & FUV & NUV & E(B-V)\\
		 & & & & km/s & & mJy & mJy & mag & mag & mag & mag & mag & mag & mag & mag \\
\hline
7ZW700 & 17h15m23.25s & 57d25'58.32" & 0.03 & 9630.00 & null & ...& 0.94 & 10.32 & 9.95 & 10.41 & 10.10 & 11.27 & 21.24 & 20.35 & 0.0364\\
ESO137006 & 16h15m3.80s & -60d54'25.8" & 0.02 & 5452.30 & -4.9 & ...& 1.35 & 7.91 & 8.12 & 7.81 & 7.02 & 8.55 & 24.70 & 19.67 & 0.1944\\
ESO269 & 12h56m40.52s & -46d55'34.3" & 0.02 & 5016.00 & -2.1 & $<$  & 6 & 10.12 & 10.19 & 7.86 & 5.78 & 10.52 & 24.70 & 19.99 & 0.1414\\
ESO3060170 & 5h40m6.69s & -40d50'12.1" & 0.04 & 10903.00 & -3.9 & $<$  & 18.2 & 9.20 & 9.06 & 8.88 & 7.63 & 10.20 & 24.70 & 18.06 & 0.0325\\
ESO351030 & 01h53m00.47s & -13d44'18.5" & 0.00 & 109.40 & -4.8 & ...& 1.35 & 7.03 & 7.11 & 6.59 & 5.81 & 7.27 & 17.03 & 15.92 & 0.0158\\
ESO428 & 7h16m31.21s & -29d19'28.8" & 0.01 & 1707.00 & -1.8 & $<$  & 81 & 8.44 & 8.10 & 4.58 & 1.67 & 8.80 & 24.70 & 17.64 & 0.1961\\
ESO443G024 & 13h1m0.80s & -32d26'29.2" & 0.02 & 5114.60 & -3 & $<$  & 1387 & 8.01 & 8.22 & 7.46 & 6.37 & 8.92 & 19.48 & 18.15 & 0.0909\\
ESO495G021 & 8h36m15.13s & -26d24'33.8" & 0.00 & 858.90 & -3.1 & $<$  & 83.8 & 8.49 & 8.18 & 3.66 & 0.34 & 9.05 & 0.00 & 0.00 & 0.1117\\
ESO552G020 & 04h54m52.26s & -18d6'56.0" & 0.03 & 9396.80 & -3.9 & ...& 1.35 & 8.89 & 8.86 & 9.57 & 10.27 & 9.54 & 21.06 & 18.65 & 0.0719\\
IC1262 & 17h33m2.02s & 43d45'34.7" & 0.03 & 9788.00 & -4.8 & $<$  & 69.1 & 9.96 & 10.19 & 10.01 & 8.48 & 10.74 & 20.31 & 18.98 & 0.0203\\
IC1459 & 22h57m10.60s & -36d27'43.9" & 0.01 & 1794.50 & -4.8 & $<$  & 1279.7 & 6.57 & 6.64 & 6.14 & 5.08 & 6.91 & 16.89 & 15.28 & 0.0162\\
IC1633 & 01h09m55.55s & -45d55'52.3" & 0.02 & 7265.20 & -3.9 & $<$  & 1.59 & 7.97 & 8.00 & 7.84 & 6.92 & 8.52 & 17.95 & 17.65 & 0.0106\\
IC1729 & 1h47m55.26s & -26d53'31.4" & 0.01 & 1503.00 & -3.8 & ...& 1.35 & 9.34 & 9.46 & 9.18 & 8.86 & 9.86 & 19.33 & 17.48 & 0.0178\\
IC4296 & 13h36m39.03s & -33d57'57.0" & 0.01 & 3781.40 & -4.9 & $<$  & 8.07 & 7.31 & 7.25 & 7.06 & 6.01 & 7.52 & 18.27 & 16.24 & 0.0618\\
IC5267 & 22h57m13.48s & -43d23'45.4" & 0.01 & 1714.90 & -1.1 & ...& 6 & 7.19 & 7.29 & 5.82 & 3.96 & 7.71 & 15.77 & 14.51 & 0.0124\\
IC5358 & 23h47m45.07s & -28d8'26.3" & 0.03 & 8651.20 & -3.9 & $<$  & 27.5 & 9.29 & 9.43 & 9.73 & 0.00 & 10.70 & 0.00 & 0.00 & 0.0188\\
NGC1023 & 2h40m24.00s & 39d3'47.7" & 0.00 & 645.00 & -2.6 & ...& 1.35 & 6.14 & 6.17 & 5.75 & 4.99 & 0.00 & 17.85 & 15.75 & 0.0605\\
NGC1052 & 2h41m4.8s & -8d15'21.0" & 0.00 & 1483.80 & -4.7 & $<$  & 1017.22 & 7.22 & 7.15 & 5.52 & 3.06 & 7.34 & 17.65 & 16.19 & 0.0266\\
NGC1132 & 02h52m51.83s & -1d16'29.1" & 0.02 & 6957.60 & -4.8 & $<$  & 5.4 & 8.67 & 8.99 & 8.71 & 6.91 & 9.20 & 19.46 & 18.90 & 0.0633\\
NGC1199 & 03h3m38.41s & -15d36'48.6" & 0.01 & 2681.60 & -4.8 & ...& 1.35 & 8.10 & 8.05 & 7.93 & 6.85 & 0.00 & 19.11 & 17.25 & 0.0559\\
NGC1265 & 3h18m15.62s & 41d51'27.8" & 0.03 & 7536.00 & -4 & $<$  & 5260 & 8.28 & 8.32 & 8.25 & 7.70 & 8.66 & 20.82 & 17.68 & 0.1623\\
NGC1316 & 03h22m41.71s & -37d12'28.7" & 0.01 & 1788.30 & -1.8 & $<$  & 119000 & 5.51 & 5.54 & 4.52 & 3.33 & 0.00 & 15.99 & 14.12 & 0.0208\\
NGC1332 & 03h26m17.2s & -21d20'7.0" & 0.01 & 1526.20 & -2.9 & $<$  & 4.6 & 6.84 & 6.90 & 6.38 & 5.64 & 7.09 & 17.01 & 16.00 & 0.0327\\
NGC1340 & 3h28m19.7s & -31d4'5.0" & 0.00 & 1183.10 & -4 & ...& 1.35 & 7.19 & 7.30 & 6.72 & 5.86 & 7.57 & 17.78 & 15.76 & 0.0185\\
NGC1381 & 3h36m31.67s & -35d17'42.7" & 0.01 & 1727.00 & -2.1 & ...& 1.35 & 8.27 & 8.33 & 7.94 & 6.99 & 8.41 & 18.61 & 16.97 & 0.0129\\
NGC1386 & 3h36m46.22s & -35d59'57.2" & 0.00 & 895.10 & -0.7 & $<$  & 37.7 & 7.85 & 7.45 & 4.41 & 1.87 & 8.16 & 17.13 & 15.87 & 0.0125\\
NGC1387 & 3h36m57.03s & -35d30'23.8" & 0.00 & 1260.50 & -2.9 & $<$  & 4 & 7.19 & 7.24 & 5.96 & 4.51 & 7.46 & 16.88 & 15.92 & 0.0125\\
NGC1389 & 03h37m11.75s & -35d44'45.9" & 0.00 & 921.00 & -2.8 & ...& 1.35 & 8.39 & 8.42 & 8.08 & 7.13 & 8.79 & 19.14 & 17.03 & 0.0114\\
NGC1395 & 03h38m29.71s & -23d01'38.7" & 0.01 & 1701.40 & -4.9 & $<$  & 1.1 & 6.70 & 6.79 & 6.27 & 5.62 & 6.94 & 16.93 & 15.67 & 0.0231\\
NGC1399 & 03h38m29.02s & -35d27'1.6" & 0.00 & 1425.70 & -4.6 & $<$  & 639 & 6.12 & 6.20 & 5.70 & 4.92 & 0.00 & 15.49 & 15.19 & 0.0125\\
NGC1404 & 03h38m51.93s & -35d35'39.0" & 0.01 & 1946.30 & -4.8 & $<$  & 3.9 & 6.58 & 6.72 & 6.07 & 5.31 & 7.00 & 17.04 & 15.71 & 0.0100\\
NGC1407 & 03h40m11.83s & -18d34'48.3" & 0.01 & 1791.40 & -4.5 & $<$  & 87.7 & 6.42 & 6.42 & 6.07 & 5.75 & 6.67 & 16.73 & 15.66 & 0.0405\\
NGC1426 & 03h42m49.10s & -22d6'30.1" & 0.00 & 1444.70 & -4.9 & ...& 1.35 & 8.43 & 8.29 & 8.10 & 7.23 & 8.70 & 18.96 & 16.70 & 0.0162\\
NGC1427 & 03h42m19.42s & -35d23'33.2" & 0.00 & 1388.20 & -4 & ...& 1.35 & 7.97 & 8.14 & 7.29 & 7.29 & 8.25 & 18.94 & 16.35 & 0.0122\\
NGC1439 & 03h44m49.95s & -21d55'14.0" & 0.01 & 1667.90 & -4.8 & ...& 1.35 & 8.31 & 8.41 & 8.06 & 6.94 & 8.60 & 19.05 & 16.75 & 0.0270\\
NGC1482 & 3h54m38.93s & -20d30'7.7" & 0.01 & 1860.40 & -0.8 & $<$  & 237.8 & 7.96 & 7.62 & 0.00 & 0.87 & 8.57 & 17.94 & 16.51 & 0.0399\\
NGC1521 & 04h8m18.93s & -21d3'6.9" & 0.01 & 4217.50 & -4.2 & $<$  & 4.2 & 8.33 & 8.16 & 8.51 & 7.46 & 9.11 & 19.90 & 17.64 & 0.0412\\
NGC1549 & 04h15m45.11s & -55d35'32.0" & 0.00 & 1243.40 & -4.3 & ...& 6 & 6.52 & 6.59 & 6.10 & 5.47 & 0.00 & 18.06 & 15.44 & 0.0124\\
NGC1550 & 04h19m37.95s & 2d24'36.0" & 0.01 & 3785.20 & -4.1 & $<$  & 21 & 8.48 & 8.54 & 8.44 & 8.19 & 8.97 & 19.11 & 18.34 & 0.1317\\
NGC1553 & 04h16m10.46s & -55d46'48.1" & 0.00 & 1148.40 & -2.3 & $<$  & 6.8 & 6.22 & 6.07 & 5.33 & 4.28 & 6.35 & 16.72 & 14.75 & 0.0151\\
NGC1587 & 04h30m39.93s & 0d39'41.8" & 0.01 & 3671.70 & -4.8 & $<$  & 130.6 & 8.28 & 8.34 & 7.90 & 7.27 & 8.70 & 19.37 & 17.96 & 0.0721\\
NGC1600 & 04h31m39.87s & -5d5'10.5" & 0.02 & 4708.20 & -4.6 & $<$  & 61.6 & 7.56 & 7.63 & 7.35 & 7.26 & 7.87 & 17.99 & 17.13 & 0.0440\\
NGC1638 & 04h41m36.5s & -1d48'32.5" & 0.01 & 3293.10 & -1.9 & ...& 1.35 & 8.97 & 8.94 & 7.76 & 7.01 & 9.69 & 18.84 & 17.52 & 0.0422\\
NGC1700 & 4h56m56.32s & -4d51'56.8" & 0.01 & 3891.40 & -4.7 & ...& 1.35 & 7.84 & 7.93 & 7.42 & 6.60 & 8.32 & 0.00 & 0.00 & 0.0433\\
NGC1705 & 4h54m13.50s & -53d21'39.8" & 0.00 & 628.30 & -2.9 & ...& 6 & 10.09 & 10.01 & 8.09 & 5.57 & 10.98 & 13.55 & 13.67 & 0.0079\\
NGC1800 & 5h6m25.55s & -31d57'13.8" & 0.00 & 800.80 & 8 & ...& 1.35 & 9.90 & 9.91 & 7.51 & 5.39 & 10.71 & 15.27 & 14.58 & 0.0143\\
NGC205 & 00h40m22.08s & 41d41'7.1" & 0.00 & -241.30 & -4.8 & ...& 1.35 & 5.94 & 6.14 & 5.57 & 4.89 & 5.90 & 15.45 & 12.95 & 0.0867\\
NGC2110 & 5h52m11.39s & -7d27'22.3" & 0.01 & 2311.70 & -3 & $<$  & 298.8 & 7.52 & 6.98 & 4.67 & 2.44 & 8.31 & 24.70 & 19.00 & 0.3730\\
NGC221 & 00h42m41.82s & 40d51'54.7" & 0.00 & -199.70 & -4.8 & ...& 1.35 & 4.91 & 4.85 & 4.17 & 3.26 & 5.13 & 16.16 & 14.24 & 0.1527\\
NGC2305 & 06h48m37.4s & -64d16'23.8" & 0.01 & 3570.00 & -4.9 & ...& 6 & 8.27 & 8.23 & 7.98 & 7.24 & 8.76 & 19.34 & 17.56 & 0.0759\\
NGC2314 & 7h10m32.5s & 75d19'36.0" & 0.01 & 3834.40 & -4.7 & $<$  & 23.7 & 8.57 & 8.74 & 8.24 & 6.67 & 9.02 & 19.56 & 18.27 & 0.0416\\
NGC2329 & 07h09m8.00s & 48d36'55.8" & 0.02 & 5792.90 & -3 & $<$  & 732 & 8.91 & 9.07 & 8.75 & 8.44 & 9.55 & 19.61 & 18.62 & 0.0721\\
NGC2340 & 07h11m10.81s & 50d10'28.8" & 0.02 & 5925.20 & -4.8 & $<$  & 0.5 & 8.55 & 8.58 & 8.42 & 8.62 & 8.89 & 19.44 & 18.25 & 0.0733\\
NGC2434 & 07h34m51.17s & -69d17'2.8" & 0.00 & 1449.70 & -4.8 & ...& 6 & 7.45 & 7.60 & 6.74 & 6.23 & 8.24 & 19.29 & 17.17 & 0.2400\\
NGC2563 & 8h20m35.69s & 21d4'4.1" & 0.02 & 4509.10 & -2.2 & ...& 0.93 & 8.47 & 8.86 & 8.67 & 8.13 & 9.16 & 19.21 & 18.37 & 0.0432\\
NGC2768 & 9h11m37.5s & 60d2'14.0" & 0.00 & 1397.80 & -4.5 & $<$  & 14.5 & 6.90 & 7.05 & 6.66 & 5.75 & 7.08 & 18.22 & 15.64 & 0.0448\\
NGC2778 & 9h12m24.37s & 35d1'39.2" & 0.01 & 2030.50 & -4.8 & ...& 0.96 & 9.34 & 9.47 & 9.05 & 7.71 & 9.56 & 19.90 & 18.02 & 0.0222\\
NGC2787 & 9h19m18.62s & 69d12'12.0" & 0.00 & 696.80 & -1 & $<$  & 10.9 & 7.07 & 7.16 & 6.42 & 5.68 & 0.00 & 18.93 & 17.07 & 0.1335\\
NGC2832 & 09h19m46.86s & 33d44'59.0" & 0.02 & 6898.90 & -4.3 & $<$  & 5.0 & 8.10 & 8.40 & 8.19 & 6.74 & 9.12 & 19.02 & 17.55 & 0.0173\\

\hline
\end{tabular}
}
Note: Column description: (1) Name of the galaxy; (2) RA in J2000; (3) DEC in J2000; (4) Redshift; (5) Radial Velocity \citep{makarov2014}; (6) Morphological type (de Voucouleur's scale) \citep{makarov2014}; (7) Limit on the radio flux at 1.4 GHz; (8) Radio flux at 1.4 GHz; (9) - (12) Photometric estimates in \wise bands; (13) Photometric estimate in \ks band; (14) GALEX FUV; (15) GALEX NUV; (16) Galactic reddening from \galex
\end{table}

\clearpage
\newpage

\begin{table}
\caption{Derived properties of a subset of the sample.}
\label{tab:derivedprop}

\begin{tabular}{lccccc}
	\toprule
	Name & SFR & $M_{*}$ & l\_P1.4 & P1.4 & $M_{K_s}$\\
	 		& $M_{\odot}yr^{-1}$ & $M_{\odot}$ &  & $WHz^{-1}$ & mag\\
\hline
\startdata
7ZW700     & 1.64E-02 & 1.17E+11 & $<$   & 1.96E+21 & -24.40\\
ESO137006  & 1.54E-03 & 4.84E+11 & $<$   & 9.00E+20 & -25.95\\
ESO269     & 8.69E-04 & 6.52E+10 & --- & 2.22E+21 & -23.77\\
ESO3060170 & 1.77E-03 & 4.00E+11 & --- & 3.32E+22 & -25.74\\
ESO351030  & 9.41E-05 & 5.97E+08 & $<$   & 3.62E+17 & -18.68\\
ESO428     & 1.53E-04 & 3.76E+10 & --- & 5.30E+21 & -23.17\\
ESO443G024 & 2.91E-02 & 2.92E+11 & --- & 8.14E+23 & -25.40\\
ESO495G021 & ---      & 7.39E+09 & --- & 1.39E+21 & -21.41\\
ESO552G020 & 5.10E-02 & 5.56E+11 & $<$   & 2.67E+21 & -26.10\\
IC1262     & 4.35E-02 & 1.96E+11 & --- & 1.49E+23 & -24.97\\
IC1459     & 2.15E-02 & 2.23E+11 & --- & 9.25E+22 & -25.10\\
IC1633     & 2.02E-01 & 8.29E+11 & --- & 1.88E+21 & -26.53\\
IC1729     & 2.65E-03 & 1.03E+10 & $<$   & 6.84E+19 & -21.77\\
IC4296     & 9.09E-03 & 5.75E+11 & --- & 2.59E+21 & -26.14\\
IC5267     & 1.25E-01 & 9.77E+10 & $<$   & 3.96E+20 & -24.21\\
IC5358     & ---      & 1.59E+11 & --- & 4.62E+22 & -24.74\\
NGC1023    & ---      & 7.78E+10 & $<$   & 1.26E+19 &    ---\\
NGC1052    & 4.69E-03 & 1.03E+11 & --- & 5.02E+22 & -24.27\\
NGC1132    & 4.91E-02 & 4.14E+11 & --- & 5.86E+21 & -25.78\\
NGC1199    & ---      & 1.18E+11 & $<$   & 2.18E+20 &    ---\\
NGC1265    & 8.21E-02 & 8.21E+11 & --- & 6.70E+24 & -26.52\\
NGC1316    & ---      & 2.71E+11 & --- & 8.54E+24 &    ---\\
NGC1332    & 2.00E-02 & 1.37E+11 & --- & 2.40E+20 & -24.58\\
NGC1340    & 2.82E-03 & 5.28E+10 & $<$   & 4.24E+19 & -23.54\\
NGC1381    & 2.40E-03 & 5.19E+10 & $<$   & 9.03E+19 & -23.52\\
NGC1386    & 8.22E-03 & 1.76E+10 & --- & 6.78E+20 & -22.35\\
NGC1387    & 1.69E-02 & 6.65E+10 & --- & 1.43E+20 & -23.79\\
NGC1389    & 2.04E-04 & 1.04E+10 & $<$   & 2.57E+19 & -21.78\\
NGC1395    & 2.09E-02 & 1.96E+11 & --- & 7.14E+19 & -24.96\\
NGC1399    & ---      & 1.76E+11 & --- & 2.91E+22 &    ---\\
NGC1404    & 1.72E-02 & 2.41E+11 & --- & 3.31E+20 & -25.19\\
NGC1407    & 3.48E-02 & 2.79E+11 & --- & 6.31E+21 & -25.35\\
NGC1426    & 1.10E-03 & 2.79E+10 & $<$   & 6.32E+19 & -22.85\\
NGC1427    & 4.95E-03 & 3.89E+10 & $<$   & 5.84E+19 & -23.21\\
NGC1439    & 9.30E-04 & 4.09E+10 & $<$   & 8.43E+19 & -23.26\\
NGC1482    & 1.95E-02 & 5.25E+10 & --- & 1.85E+22 & -23.54\\
NGC1521    & 2.37E-02 & 1.63E+11 & --- & 1.68E+21 & -24.77\\
NGC1549    & ---      & 1.48E+11 & $<$   & 1.37E+20 &    ---\\
NGC1550    & 5.48E-02 & 1.55E+11 & --- & 6.75E+21 & -24.71\\
NGC1553    & 2.88E-03 & 1.53E+11 & --- & 1.38E+20 & -24.70\\
NGC1587    & 8.43E-03 & 1.83E+11 & --- & 3.95E+22 & -24.89\\
NGC1600    & 7.29E-02 & 6.44E+11 & --- & 3.06E+22 & -26.26\\
NGC1638    & 2.95E-02 & 5.88E+10 & $<$   & 3.28E+20 & -23.66\\
NGC1700    & ---      & 2.89E+11 & $<$   & 4.59E+20 & -25.39\\
NGC1705    & 1.40E-01 & 6.41E+08 & $<$   & 3.49E+19 & -18.75\\
NGC1800    & 4.89E-02 & 1.34E+09 & $<$   & 1.94E+19 & -19.55\\
NGC205     & 5.00E-03 & 1.05E+10 & $<$   & 1.76E+18 & -21.79\\
NGC2110    & 1.10E-03 & 1.14E+11 & --- & 3.58E+22 & -24.38\\
NGC221     & 1.63E-03 & 1.48E+10 & $<$   & 1.21E+18 & -22.16\\
NGC2305    & 1.14E-02 & 1.64E+11 & $<$   & 1.13E+21 & -24.77\\
NGC2314    & 3.65E-03 & 1.47E+11 & --- & 7.82E+21 & -24.66\\
NGC2329    & 4.11E-02 & 2.09E+11 & --- & 5.51E+23 & -25.03\\
NGC2340    & 2.81E-02 & 4.01E+11 & --- & 3.94E+20 & -25.74\\
NGC2434    & 1.55E-02 & 4.59E+10 & $<$   & 1.86E+20 & -23.39\\
NGC2563    & 2.37E-02 & 1.79E+11 & $<$   & 4.24E+20 & -24.87\\
NGC2768    & 1.25E-02 & 1.17E+11 & --- & 6.36E+20 & -24.40\\
NGC2778    & 8.63E-04 & 2.50E+10 & $<$   & 8.88E+19 & -22.73\\
NGC2787    & ---      & 1.30E+10 & --- & 1.19E+20 &    ---\\
NGC2832    & 5.00E-02 & 4.32E+11 & --- & 5.34E+21 & -25.83\\

\hline
\end{tabular}

Note: Column description: (1) Name of the galaxy; (2) Star Formation Rate; (3) Stellar Mass; (4) Limit on the radio flux at 1.4 GHz; (5) Radio power at 1.4 GHz; (6) Absolute \ks band magnitude
\end{table}

\clearpage
\newpage

\appendices
\appendixpage
\setcounter{table}{0}
\renewcommand{\thetable}{A\arabic{table}}

\section{Calculations}
	From the apparent magnitude, absolute magnitude is calculated using the following distance modulus relation:
  \begin{equation}
    M_{k} = m_{k}-5logD_{L} + 5 \;,
  \end{equation}

  where, $m_{k}$ is the apparent magnitude in \ks band, distance $D_{L}$ is the luminosity distance calculated using the relation
  \begin{equation}
   D_L = v/H_o\;,
   \label{eq:distance}
  \end{equation}
  where $v$, the radial velocity is obtained from hyperleda \footnote{http://leda.univ-lyon1.fr}, and $H_o$ is the Hubble constant.\\
Redshift is calculated using the radial velocity with the relation $z=v/c$. \\

 Flux densities for $K_{s}$  and \emph{WISE} bands in the units of $W/cm^{2}/\mu m$ are calculated using the zero-magnitude fluxes given in Table ~\ref{tab:zeromagflux}.

  \begin{table}[!h]
    \caption{\twomass \citep{cohen2003} and \emph{WISE} \citep{jarrett2011} fluxes for zero-magnitude.}
    \label{tab:zeromagflux}
      \begin{center}
      \begin{tabular}{ccc}
	\hline
	{Band} & {$F_{\lambda}-0\;mag\;(W/cm^{2}/\mu m$)} & {$F_{\nu}-0\;mag\;(Jy)$}\\
	\hline
	  {$K_{s}$} & {4.283E-14} & {666.7}\\
	  {W1} & {8.1787E-15} & {309.540}\\
	  {W2} & {2.4150E-15} & {171.787}\\
	  {W3} & {6.5151E-17} & {31.674}\\
	  {W4} & {5.0901E-18} & {8.363}\\
	\hline\\
      \end{tabular}
      \end{center}
  \end{table}

\section{Photometry comparisions}

		\subsection{Comparision with the 2MASS}
	A comparison of our \ks  band measurements with that of the 2MASS XSC catalog has been done.  We compare our results with the isophotal measurements that are set to 20 mag \sqarcsec surface brightness isophote at \ks using elliptical apertures (identified with k\_m\_k20fe in the catalog).  This corresponds to roughly 1$\sigma$ of the typical background noise in the \ks images. \\
	
	The 2MASS ellipse fitting pipeline is described in \citep{jarrett2000}.   The basic step in the fitting method is to first isolate an approximate 3$\sigma$ isophote which at \ks band magnitude is at 18.55.  This step is called \lq isovector operation\rq and is done by analyzing the radial profiles at different position angles and determining the pixels that correspond to the 3$\sigma$isophote.  The center of the isophote corresponds to the pixel with peak intensity.  A best-fit ellipse to the pixel distribution is obtained by minimizing the ratio of the standard deviation to the mean of the radial distribution at the 3$\sigma$ isophote.  Using the axis ratio and position angle of the 3$\sigma$ isophote, an ellipse fit to \ks band at 20 mag \sqarcsec is obtained by varying the semi-major axis such that the mean surface brightness along the ellipse is 20 mag \sqarcsec.  The integrated flux within this ellipse after background subtraction is the 2MASS isophotal magnitude.  \\
	
\begin{figure}
	\centering
		\includegraphics[width=0.5\textwidth]{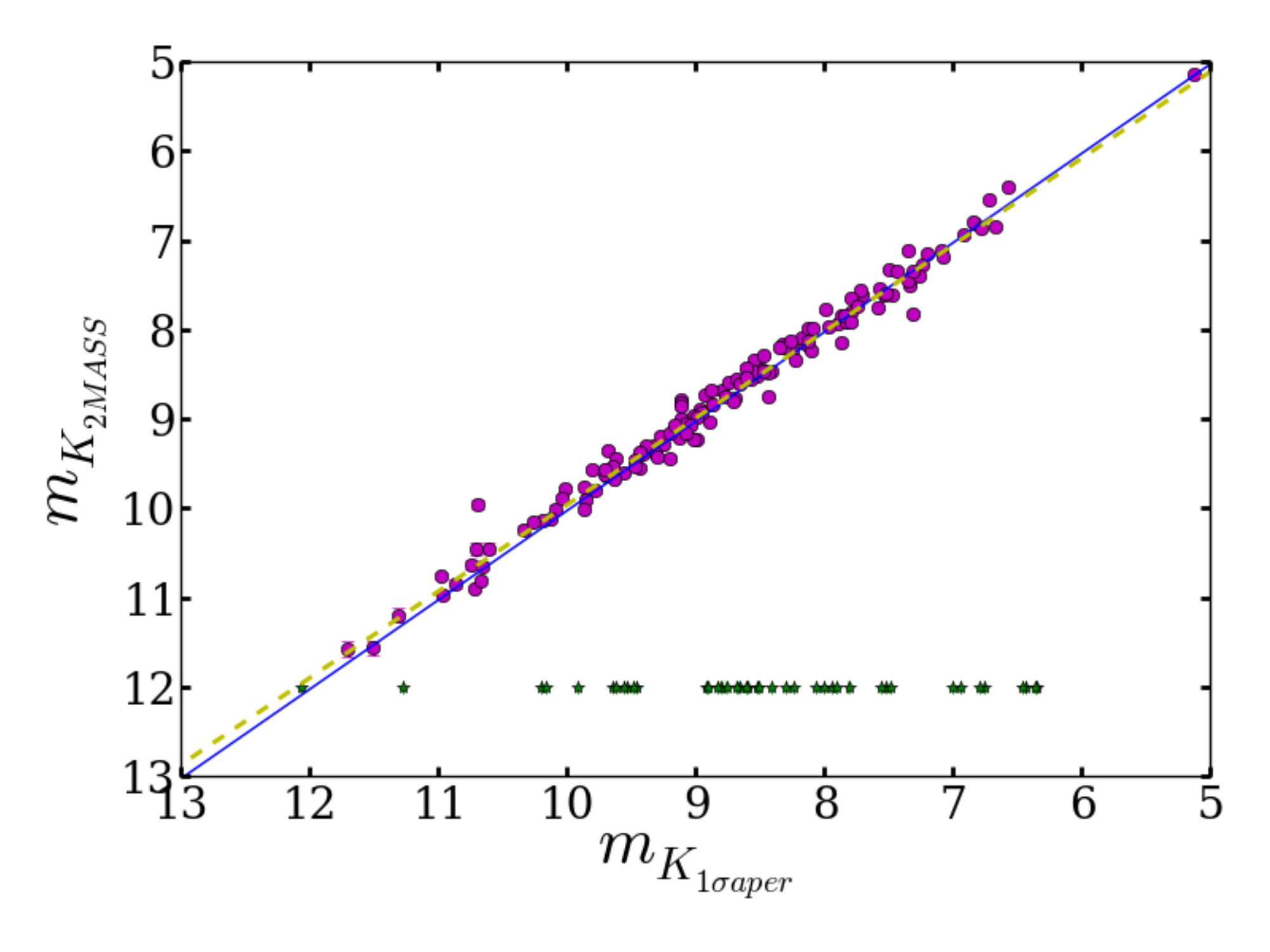}
		\caption{Comparison between 2MASS \ks magnitude from the catalog ($m_{K_{2MASS}}$ ) and our measurements ($m_{K_{1\sigma aper}}$).  Solid blue line shows the one-to-one relation between the x and y.  The yellow dashed line is the linear fit with a slope of 0.97 and an intercept of 0.23.  Our magnitude measurements match with that of the 2MASS magnitude.  The green stars are the galaxies whose 2MASS magnitudes are not present.  The outlier galaxy is IC5358 whose 2AMSS mag is brighter than our measurement.  This is because of the presence of a companion galaxy which was not excluded in the catalog estimates.}  
		\label{fig:k}
	\end{figure}

	The Figure ~\ref{fig:k} shows the comparison with the 2MASS.  Our magnitude measurements match closely with that of 2MASS catalog.   Galaxies that do not have 2MASS magnitudes are shown with green stars.  Out of 231 galaxies in the sample, estimates for 47 galaxies were excluded, either because they are close to the edge of the image or due to the presence of a very close companion which could not be masked.

	\subsection{Comparison with WISE}
	The WISE extended source photometry pipeline uses the aperture that is based on the elliptical shape reported by the 2MASS XSC.  Due to the larger beam size of WISE, the aperture is scaled accordingly.  They sum the pixel fluxes within this aperture and subtract it with the background to obtain the elliptical aperture  photometry measurement which is indicated by w?gmag in their catalog.  Since W1 is the most sensitive wavelength for all galaxies where the emission is from the evolved stellar population, the 2MASS \ks aperture is typically 3 to 4 times smaller than the 1$\sigma$ aperture for W1.  This underestimates the integrated flux by about 30-40\%.  Hence, we performed photometry on the WISE images to estimate the galaxy magnitude within an elliptical aperture fit to 1$\sigma$ isophote.\\
	
\begin{figure}
	\centering
		\includegraphics[width=0.5\textwidth]{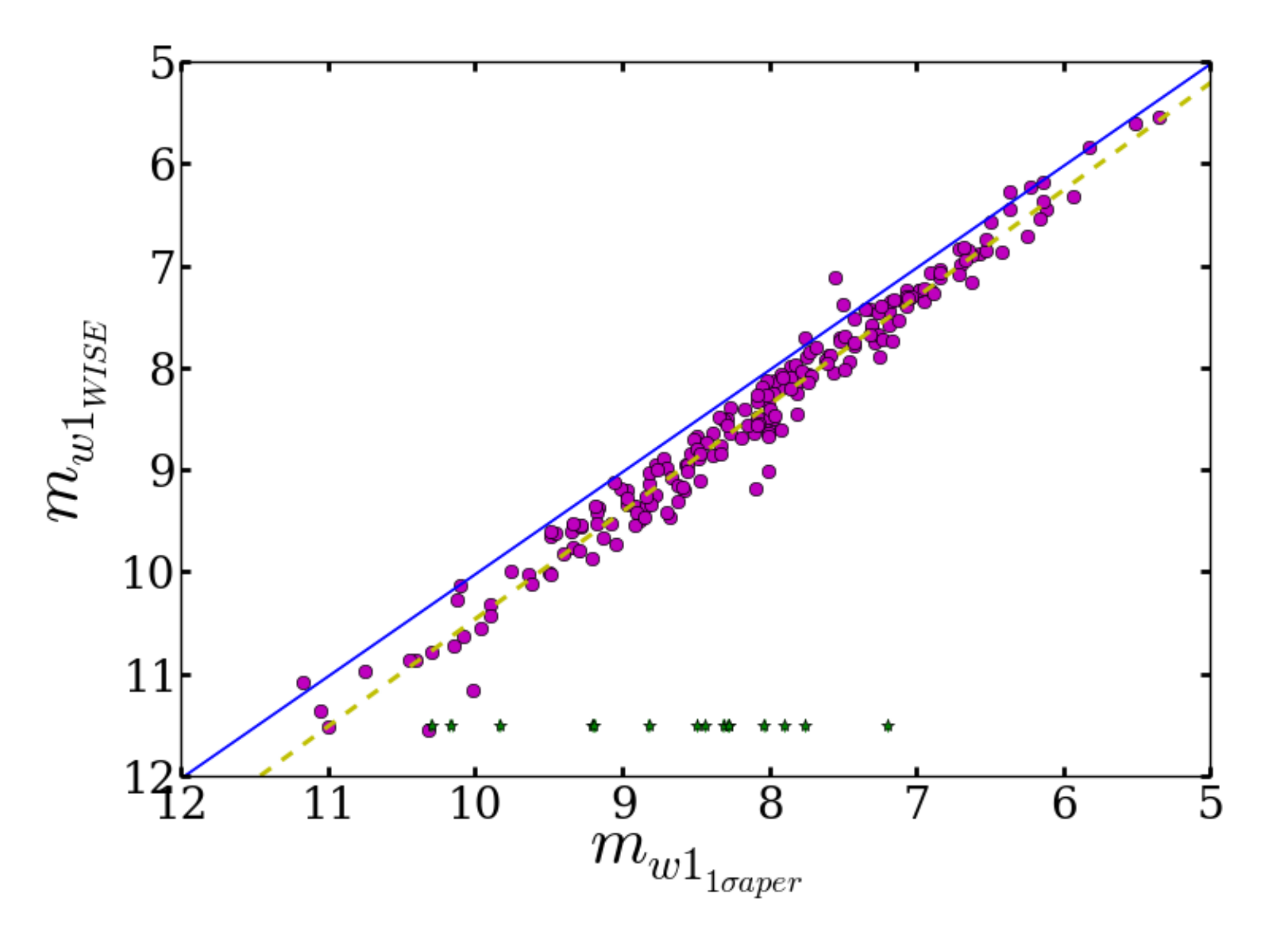}
		\caption{Comparison between WISE W1 magnitude with our measurements.  Notation used in this figure is same as that used in Figure ~\ref{fig:k}.  The WISE catalog measurements give galaxy magnitudes that are faint compared to the magnitudes that we measured.  The uncertainty in magnitude is smaller than the point size.  The mean difference between WISE catalog and our measurements is $\sim$0.34 mag with a range of percentage difference between  $\sim$5\% to 70\%.  The slope of the fit is 1.05 with an intercept of $-0.06$} 
		\label{fig:w1}
	\end{figure}

	Shown in Figure ~\ref{fig:w1} is the comparison with WISE W1 magnitude.  The WISE W1 catalog magnitudes are faint compared to our measurements on an average by 0.31 mag.   There are few outliers whose WISE magnitudes are brighter to our measurements.  NGC4467 at mag of ~11 is brighter by 0.1 mag.  This is because the WISE aperture is bigger than the size of the galaxy because of which some of the light from the adjacent galaxy is also included.  In case of NGC4782, NGC5353 NGC821, the emission from the nearby source increased the WISE W1 flux.  These are consistently brighter in W2 band as well.\\
	
\begin{figure}
	\centering
		\includegraphics[width=0.5\textwidth]{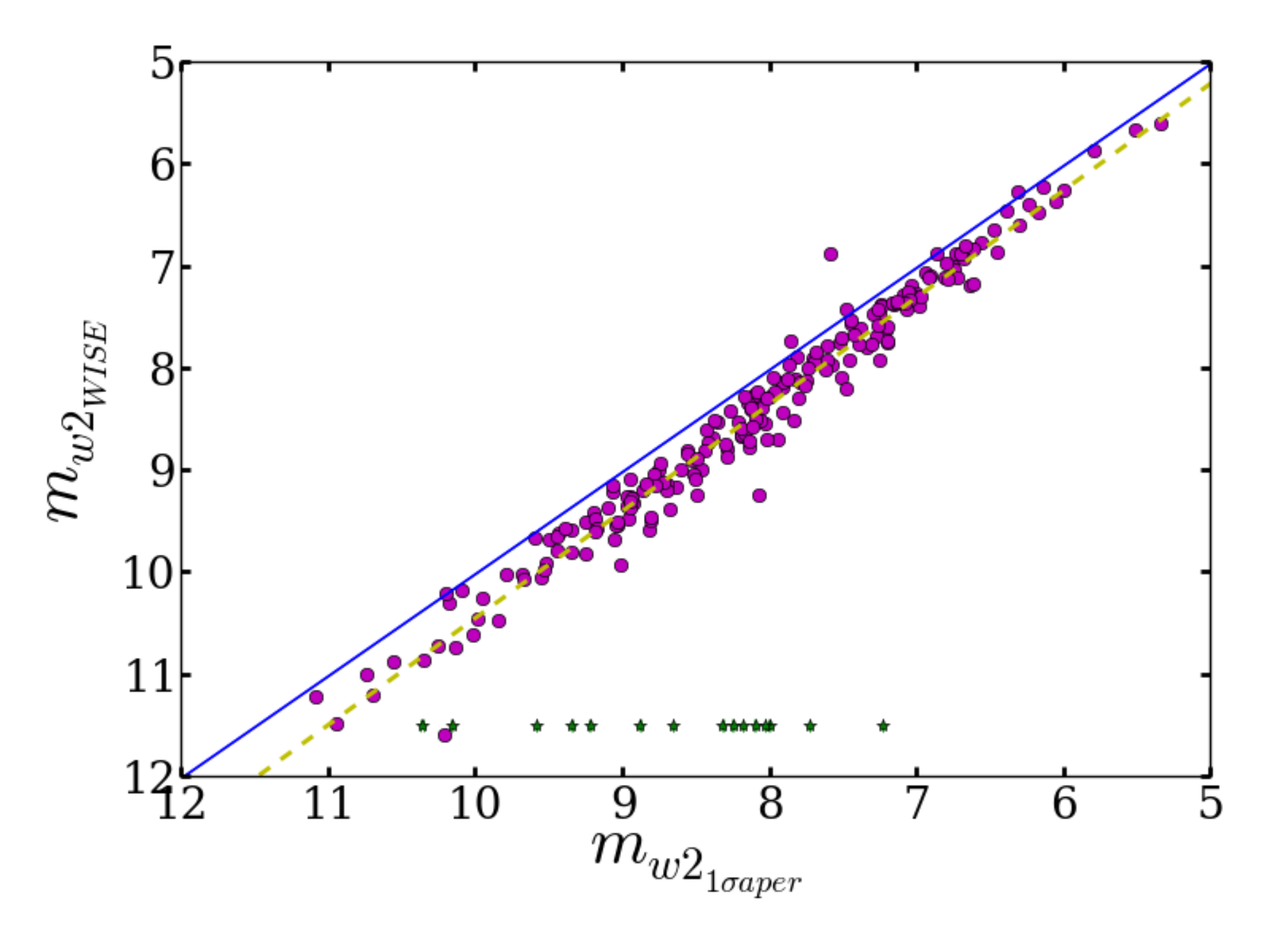}
		\caption{Comparison between WISE W2 magnitude with our measurements.  The WISE magnitudes are fainter by ~0.32 mag.  the slope of the fit is 1.04 with an intercept of $-0.07$. }
		\label{fig:w2}
	\end{figure}

\begin{figure}
	\centering
		\includegraphics[width=0.5\textwidth]{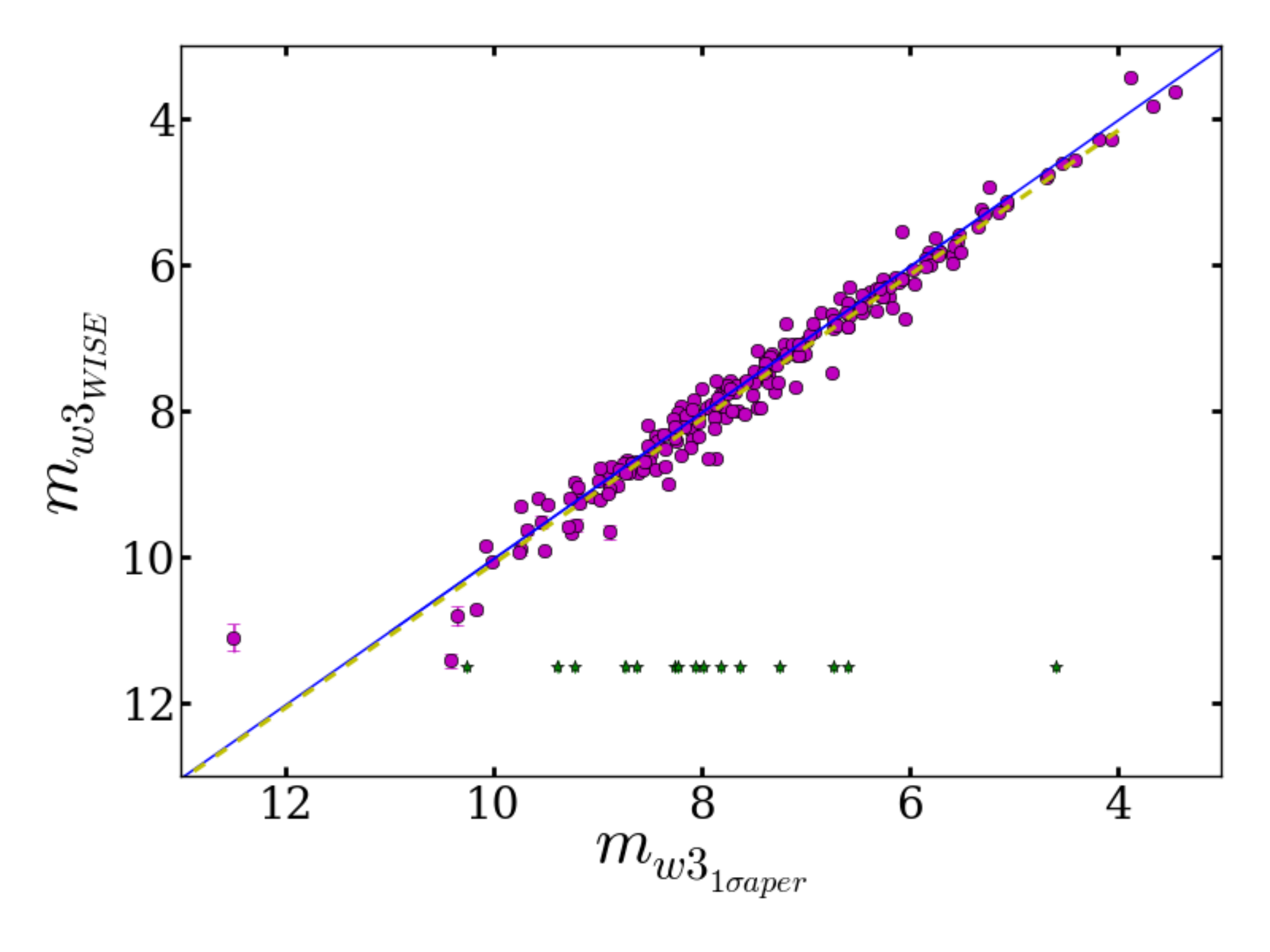}
		\caption{Comparison between W3 WISE magnitude with our measurements.  It can be noticed that the WISE catalog measurements give galaxy magnitudes that are close to the magnitudes that we measured.  The mean difference is $\sim$0.095.}  
		\label{fig:w3}
	\end{figure}

\begin{figure}
	\centering
		\includegraphics[width=0.5\textwidth]{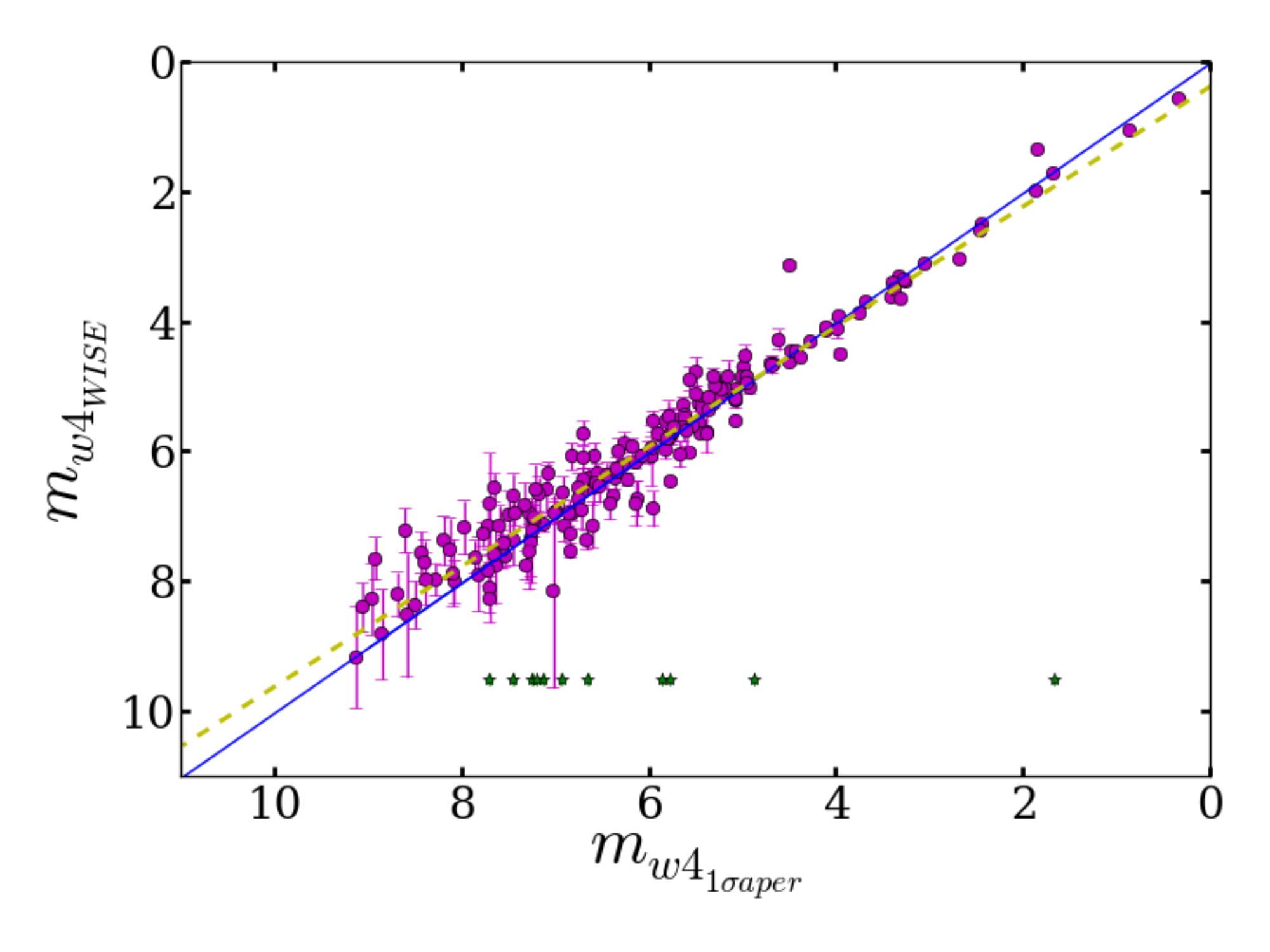}
		\caption{Comparison between W4 WISE magnitude with our measurements.  There is huge scatter in the measurements for faint galaxies and the catalog estimates small magnitudes for the faint galaxies.  The slope of the fit is 0.92 with an intercept of 0.39.} 
		\label{fig:w4}
	\end{figure}

Figures ~\ref{fig:w2} through ~\ref{fig:w4} show the comparison of W2, W3 and W4 bands.  The WISE W2 magnitudes also are fainter than our measured values.  In the case of W3 band, the extended emission from the galaxy is contained within the WISE apertures.  Hence we see that the one-to-one correspondence matches with the line fit for W3 magnitude.  \\

We modified our program when we measure W3 and W4 magnitudes.  We do not go through the two stage process in these cases and ignore the process of obtaining the 3$\sigma$ aperture and fixing the ELLIPSE parameters.  This is because, the flux from the galaxy is very faint and is almost always about 3$\sigma$ or less.  \\

Figure ~\ref{fig:w4} shows the W4 magnitude comparison.  Majority of the galaxies have very faint emission at this wavelength.  Since the apertures used in WISE measurements are larger, the magnitude measurements indicate them as bright galaxies due to contamination from faint foreground stars that are not subtracted in the images.  In our method, we use an aperture at 1$\sigma$ which gives an aperture that is just right to measure the galaxy flux without adding noise.\\

\begin{figure}
	\centering
		\includegraphics[width=0.5\textwidth]{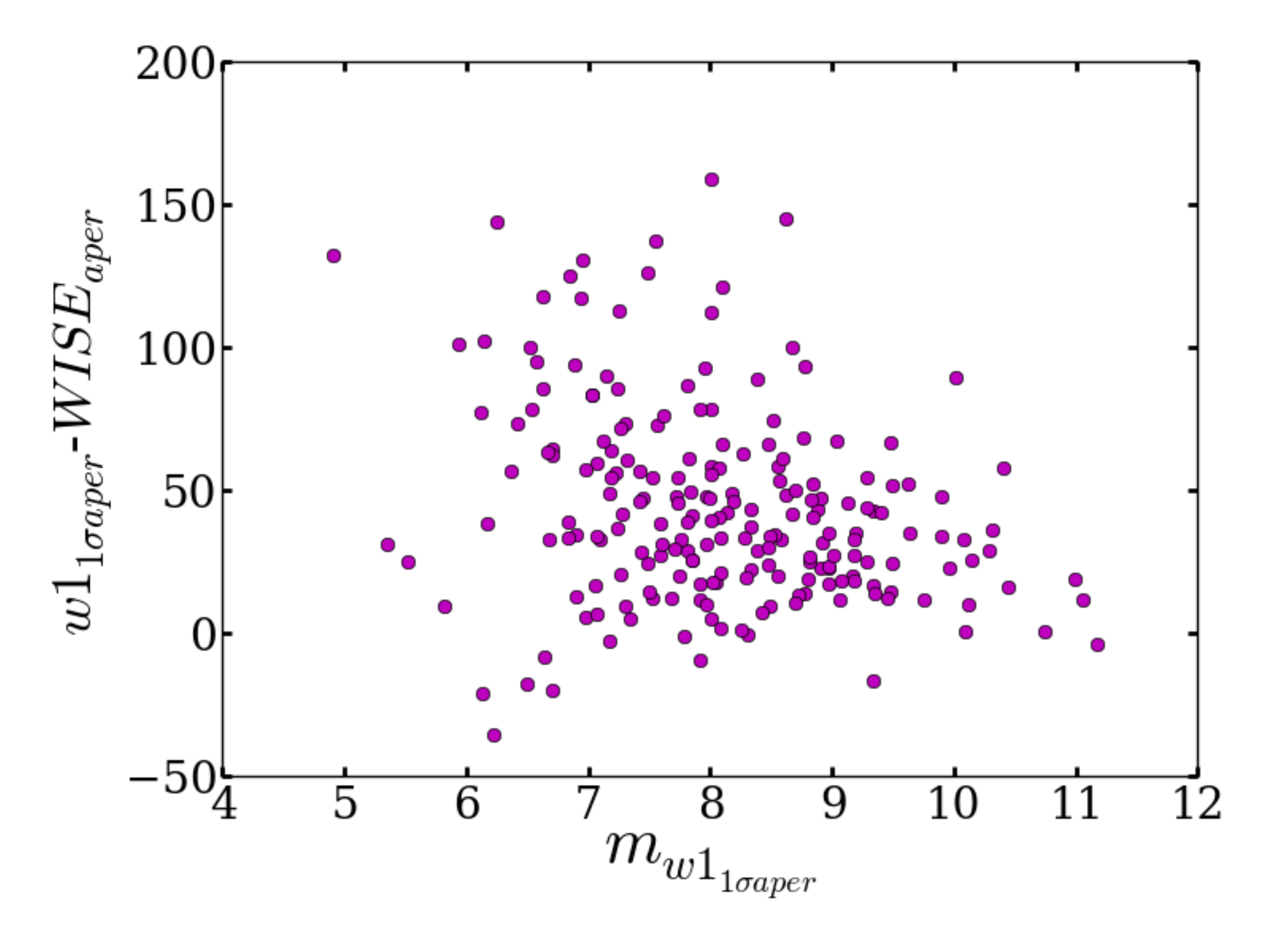}
		\caption{Difference between the aperture sizes obtained using our method and that used in the WISE catalog.}
		\label{fig:smadiff}
	\end{figure}

	We show the difference in the aperture between our method and that from WISE in W1 mag in Figure ~\ref{fig:smadiff}.  About 95\% of the galaxies have small WISE apertures.  

	\subsection{Error Analysis}
		The uncertainty in the flux measurement include contribution from possion noise  and error in the sky background estimation.   The possion noise due to the photoelectrons collected by the CCD have the following relationship with the signal.\\
		\begin{equation}
		\sigma_{source} = \sqrt{flux}
		\end{equation}
		Uncertainity in the source flux and sky flux is added in quadrature to estimate the total uncertainty.\\
		
		\begin{equation}
		\sigma_{tot} = \sqrt{{\sigma_{source}}^2 + {\sigma_{sky}}^2}
		\end{equation}
\\

\bibliographystyle{apj}	

\end{document}